\documentclass[prb,twocolumn,amsmath,amssymb,superscriptaddress]{revtex4}
\usepackage{graphicx}
\usepackage{dcolumn}
\usepackage{bm}

\newcount\bozza \bozza=1
\ifnum\bozza=1
\newdimen\shift \shift=-2truecm
\def\lb#1{%
{\label{#1}\rlap{\kern\shift{$\scriptstyle#1$}}}}
\else\def\lb#1{\label{#1}} \fi

\begin{document}



\title{Charged-phonon theory and Fano effect in the optical spectroscopy
of bilayer graphene}

\author{E. Cappelluti}
\affiliation{Instituto de Ciencia de Materiales de Madrid,
CSIC, Madrid, Spain}
\affiliation{Institute for Complex Systems, U.O.S. Sapienza, CNR, Roma, Italy}

\author{L. Benfatto}
\affiliation{Institute for Complex Systems, U.O.S. Sapienza, CNR, Roma, Italy}
\affiliation{Dipartimento di Fisica, Universit\`a ``La Sapienza'', Roma, Italy}

\author{M. Manzardo}
\affiliation{Institute for Theoretical Solid State Physics, 
IFW Dresden, Helmholtzstr. 20, 01069 Dresden, Germany}

\author{A.B. Kuzmenko}
\affiliation{DPMC,
Universit\'{e} de Gen\`{e}ve, 1211 Gen\`{e}ve,
Switzerland}

\begin{abstract}
Since their discovery,
graphene-based systems represent an exceptional
playground to explore the emergence of peculiar quantum effects.
The present paper focuses on the
anomalous appearence of strong infrared phonon resonances in the
optical spectroscopy of bilayer graphene and on their pronounced 
Fano-like asymmetry, both tunable in gated devices. By developing a
full microscopic many-body approach for the optical phonon response
we explain how both effects can be quantitatively accounted for by the
quantum interference of electronic and phononic excitations.  
We show that the phonon modes borrow a large dipole intensity from the
electronic background, the so-called charged-phonon effect, and at
the same time interfer with it, leading to a typical Fano
response. Our approach allows one to disentangle the correct selection
rules that control the relative importance of the two
(symmetric and antisymmetric) relevant phonon
modes for different values of the doping and/or of the gap in bilayer
graphene. Finally, we discuss the extension of the same theoretical
scheme to the Raman spectroscopy, to explain the lack of the same
features on the Raman phononic spectra. Besides its remarkable success
in explaining the existing experimental data
in graphene-based systems, the
present theoretical approach offers a general scheme for the
microscopic understanding of Fano-like features in  a
wide variety of other systems.
\end{abstract}

\date{\today}


\maketitle

\section{Introduction}

The peculiar properties of single and multilayer graphenes
make these systems the promising basis for the
future generation of electronic devices.
Within this context,
the analysis of the spectral properties of the phonon anomalies
observed by means of different optical probes has provided a powerful
tool not only for the characterization of the samples
but also for the investigation of the underlying scattering mechanisms
related to the electron-lattice interaction.
Large part of the investigation along this line has been based on the Raman
spectroscopy.\cite{ferrari,yan1,pisana,das1,malard,yan2,basko1,calizo,casiraghi,chen}
Typical main features under investigation within this context
were the frequency and the linewidth of the phonon anomalies,
whose trend as a function of doping was found to
be in good agreement with what expected from the theoretical
calculations of the phonon self-energy.\cite{ando1L,ando2L}
As an alternative route, phonon peak anomalies at $\omega \approx 0.2$ eV
were detected
also in the mid-infrared optical conductivity of bilayer graphene.\cite{noi,tang}
Quite interestingly, unlike in the Raman spectroscopy,
in this case a strong dependence of the phonon peak intensity
as well as of its lineshape asymmetry on the gate voltage was
reported. Understanding and controlling the 
mechanisms responsible for these features 
at relatively small doping and perpendicular electric fields
is of fundamental importance.

\begin{figure}[b]
\begin{center}
\includegraphics[scale=0.28,clip=]{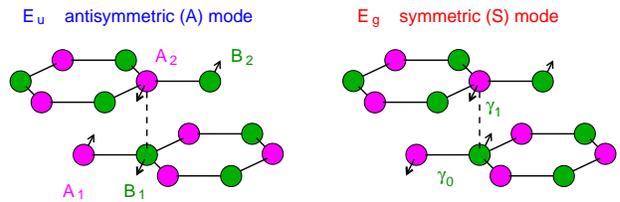}
\end{center}
\caption{Atomic structure of bilayer graphene
and lattice displacements for the $E_u$ antisymmetric (A) phonon mode
and  for the $E_g$ symmetric (S) mode.
Labels denote the A and B sublattice in each plane.
Solid links between atoms represent the in-plane $\gamma_0$ hopping,
vertical dashed links the interplane hopping $\gamma_1$.
}
\label{f-basis}
\end{figure}

From a theoretical point of view, the very evidence
of a strong infrared (IR) phonon activity, as the one reported in
Refs. \onlinecite{noi,tang}, can be considered
puzzling in graphenes which, having atoms of only one specie (carbon),
present a very small intrinsic dipole.
In bilayer graphene, for instance,
there are four carbon atoms in the unit cell,
as depicted in Fig. \ref{f-basis}, where
atoms B$_1$ and A$_2$ are connected by the vertical hopping $\gamma_1$.
There are two in-plane optical modes:
an infrared active anti symmetric (A) $E_u$ mode, which corresponds to
(out-of-phase) lattice displacements in the two layers (Fig. \ref{f-basis}a);
and a symmetric (S) $E_g$ mode, which is associated with
in-phase displacements and is Raman active (Fig. \ref{f-basis}b).
If the two layers were completely decoupled, all the atoms
would be exactly equivalent to each other.
It is thus clear by direct inspection that,
although allowed by symmetry,
the $E_u$ mode would not induce any dipole and hence
it would have no IR intensity.
Beyond this simple model of decoupled layers,
in real systems,
the interlayer hopping would induce a
slight inequivalence between the atoms (A$_1$, B$_2$) and
atoms (A$_2$, B$_1$) and hence a finite electrical dipole under the $E_u$
lattice displacements. The static dipole associated with such
physics is however three orders of magnitude smaller than what is experimentally
observed \cite{noi}, so that this effect alone cannot account for the huge
increase of the phonon intensity upon gate-induced doping
reported in Ref. \onlinecite{noi}.

A guideline to understand the origin of this huge enhancement
comes from the comparison with other carbon-based compounds,
like fullerenes, where also a similar increase of the phonon
intensity upon (chemical) doping was observed.
This effect was explained in those materials in terms
of a charged-phonon theory.\cite{rice,rice2}

The basic idea of the charged-phonon model can be understood
by considering the electronic current ($j$) response function
\begin{eqnarray}
\chi_{jj}(\omega)
&=&
-\int dt 
\langle
T_t j(t)j(0)\rangle\exp[i\omega t],
\label{chijj}
\end{eqnarray}
which is related, within the Kubo approach, to the optical conductivity as:
\begin{eqnarray}
\sigma(\omega)
&\approx&
- \frac{\chi_{jj}(\omega)}{i\hbar \omega}.
\end{eqnarray}

\begin{figure}[t]
\begin{center}
\includegraphics[scale=0.28,clip=]{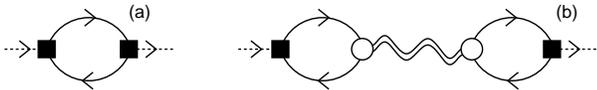}
\end{center}
\caption{(a) Lowest order contribution to the
current-current response function. (b)
Diagrams involved in the charged-phonon effect.
Solid and wavy lines are electronic and phonon propagators,
respectively,
white circles are the electron-phonon scattering operator
while black squares represent here the current operator that couples
light (dashed lines) to the particle-hole excitations.}
\label{f-diagrams}
\end{figure}

A typical diagram contributing to the electronic background
at the lowest order is depicted in Fig. \ref{f-diagrams}a,
corresponding to the single-bubble approximation.
Effects of the electron-phonon interaction {\em on the
electronic background} are commonly taken into account
by replacing the non-interacting Green's functions in Fig. \ref{f-diagrams}a
with the Green's functions evaluated in the presence of
electron-phonon interaction.\cite{stauber,carbotte_optep}
These processes lead to a
smearing of the optical features and to a redistribution of the optical
spectral weight. They are however not associated with the onset
of resonances at the characteristic phonon energies
in the optical conductivity.

A most interesting class of diagrams, analyzed by M.J. Rice in the
context of the charged-phonon effect, is depicted in Fig. \ref{f-diagrams}b.
Their contribution to the resulting optical conductivity
can be described as:\cite{rice,rice2}
\begin{eqnarray}
\Delta\chi_{jj}(\omega)
&=&
\left|\chi_{j {\rm ph}}(\omega)\right|^2D_{\rm ph}(\omega),
\end{eqnarray}
where $D_{\rm ph}(\omega)$ is the phonon propagator
of the IR active phonon mode considered,
and $\chi_{j {\rm ph}}(\omega)$ is the ``mixed'' response
function between the current and the electron-phonon
scattering operators (see Sec. \ref{s:rice} for a more detailed definition).
The resulting optical conductivity of this contribution is thus
proportional to the phonon propagator of the optically-coupled
lattice vibrations, and it presents typical
resonances at the corresponding phonon frequencies. 

It is worth stressing that 
the phonon becomes here optically visible thanks to an
intermediate process [$\chi_{j {\rm ph}}(\omega)$]
where the light couples to particle-hole electronic
excitations.
The function
$\chi_{j {\rm ph}}$ acts therefore as a prefactor of the
magnitude of the electronically-induced phonon resonances.
Pristine fullerenes and organic materials,
for which the charged-phonon theory was originally proposed,\cite{rice,rice2}
in the absence of doping are semiconductors with
a band gap significantly larger than the phonon energies.
In this case the complex function $\chi_{j {\rm ph}}$ can be
reasonably assumed to be real
and proportional to
the charge  concentrations $n$,
\begin{eqnarray}
\Delta\chi_{jj}(\omega)
&\propto&
n
D_{\rm ph}(\omega).
\label{cpt}
\end{eqnarray}

Eq. (\ref{cpt}) summarizes in a nut-shell the essence of the
charged-phonon effect, where the infrared phonon activity is triggered-in by the coupling 
of a lattice mode $\nu$ with the optically-allowed
electronic particle-hole excitations.\cite{rice,rice2}
As it was shown in Ref. \onlinecite{cbk}, the physics underlying the
charged-phonon effects is intimately
related to the onset of Fano-like
lineshape asymmetries.\cite{fano}
Such unified charged-phonon-Fano theory was also employed
to analyze the spectral properties of infrared optical phonon
in pristine graphite\cite{manzardo} and in multilayer graphenes with different
stacking orders.\cite{li}

The purpose of the present paper is to provide a detailed
microscopic derivation of the charged-phonon effect
in the optical spectroscopy of graphenes.
To this aim we focus on bilayer graphene as
the simplest and paradigmatical example.
We will show how all the information related to
the phonon intensity and Fano asymmetry can be evaluated
in terms of a unique quantity: the
current/electron-phonon response function $\chi_{j\nu}$.
Within this context we evaluate
the dependence of the optical properties of the phonon resonance
on microscopical parameters tunable by means of external gating.
The correspondence between the Fano theory
and the charged-phonon effect is derived microscopically,
and a generalization of the charged-phonon effect to the Raman
response is also provided.

The structure of the paper is the following.
In Section \ref{s:rice} we summarize the main concepts
of the charged-phonon theory and introduce
the mathematical tools employed.
In Section \ref{s:fano} we present an analytical discussion
about the correspondence between the Fano
and the charged-phonon theories,
and a suitable quantification of the optical properties
of the phonon resonances is introduced.
The detailed evaluation of the infrared properties
in the specific case of ungapped bilayer graphene is presented
in Section \ref{s-nogap}, which is generalized in Section \ref{s-gap}
to the case of gapped bilayer graphene in presence
of external gate voltage.
The role of the electronic structure and of the breaking
of particle-hole symmetry in the band structure
is discussed in Section \ref{s-TB} in bilayer graphene
and compared with bulk graphite.
A generalization of the charged-phonon theory and
of the Fano interference analysis for the Raman spectroscopy,
within an effective mass approximation, is finally presented
in Section \ref{s-raman}.
A summary of the present work and conclusion
can be found in Section \ref{s-conclusions}.
Appendices \ref{app-chijA} and \ref{app-raman}
provide all the details of the analytical evaluation of the charge-phonon theory
for IR spectroscopy of the ungapped bilayer graphene in the clean
limit, and details about a suitable generalization for the Raman response.

\section{Charged-phonon theory in bilayer graphene}
\label{s:rice}

In order to apply at a quantitative level the concepts
of the charged-phonon theory in graphenes,
in this Section
we introduce the electronic band structure and the electron-phonon
Hamiltonians as well as the relevant response functions which will
provide the analytical tools to investigate the properties of
the phonon peaks in the optical conductivity and in the Raman
response.
We focus here on the bilayer system as the most simple and representative
since the single-layer graphene does not present any IR phonon mode.
We work
in the $4 \times 4$ basis of the atomic orbitals,
as depicted in Fig. \ref{f-basis}a.
We introduce the four-vector defined as
$\Psi^\dagger_{{\bf k},\sigma}=
(a_{1{\bf k},\sigma}^\dagger,b_{1{\bf k},\sigma}^\dagger,
a_{2{\bf k}\sigma}^\dagger, b_{2{\bf k},\sigma}^\dagger)$,
where $a_{i{\bf k},\sigma}^\dagger$ and
$b_{i{\bf k},\sigma}^\dagger$
operators create an electron with spin $\sigma$ in the layer $i$ and on the
sublattice A or B, respectively.

Considering for simplicity a simple $\gamma_0$-$\gamma_1$ model,
and including a possible asymmetry between the upper and lower layer
 induced by a gate voltage,
we can write the non-interacting electronic Hamiltonian as:
\begin{eqnarray}
\label{h0}
H_0
&=&
\sum_{{\bf p},\sigma}
\Psi_{{\bf p},\sigma}^\dagger
\hat{H}_{\bf p}
\Psi_{{\bf p},\sigma},
\end{eqnarray}
where
\begin{eqnarray}
\hat{H}_{\bf p}
&=&
\left(
\begin{array}{cccc}
\Delta/2 & \gamma_0 f_{\bf p}
& 0 & 0\\
\gamma_0 f_{\bf p}^*
& \Delta/2 & \gamma_1 & 0 \\
0 & \gamma_1
& -\Delta/2 & \gamma_0 f_{\bf p} \\
0 & 0
& \gamma_0 f_{\bf p}^* &  -\Delta/2
\end{array}
\right),
\end{eqnarray}
$\gamma_0$, $\gamma_1$ are the nearest neighbor
in-plane and interplane tight-binding hopping parameters,
respectively,
and
$f_{\bf p}=\mbox{e}^{-ip_xa/\sqrt{3}}+2\mbox{e}^{ip_xa/2\sqrt{3}}\cos(p_ya/2)$.

\begin{figure}[t]
\begin{center}
 \includegraphics[scale=0.35,clip=]{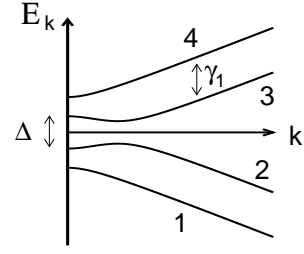}
\end{center}
\caption{Scheme of the band structure close to the K point.
The interlayer hopping parameter $\gamma_1$
determines the splitting at high energy of the 3 and 4 bands.
The interlayer different potential $\Delta$ determines
the size of the energy gap at the K (${\bf k}=0$) point.
}
\label{f-bands}
\end{figure}

Close to the K=$(4\pi/3a,0)$ point,
writing ${\bf p}=\mbox{K}+{\bf k}$, and
linearizing around the K point, we can also write:
\begin{eqnarray}
\label{hamiltonian}
\hat{H}_{\bf k}=
\left(
\begin{array}{cccc}
\Delta/2 & v {\bf \pi}_- & 0 & 0\\
 v {\bf \pi}_+ & \Delta/2 & \gamma_1  & 0\\
0 & \gamma_1 & -\Delta/2 & v {\bf \pi}_- \\
0 & 0 & v {\bf \pi}_+ & -\Delta/2
\end{array}
\right).
\end{eqnarray}
Here
$v= 10^6$ m/s is the Fermi velocity for single-layer graphene,
$\gamma_1=0.39$ eV is the interlayer hopping, $\Delta$ is the
electrostatic energy difference between the layers and
$\pi_\pm=\hbar(k_x \pm i k_y)$.  The electronic Green's function in the imaginary space
is thus expressed as a $4 \times 4$ matrix,
$\hat{G}({\bf k},i\omega_n)=
1/[(i\hbar\omega_n+\mu)\hat{I}-\hat{H}_{\bf k}]$,
where $\mu$ is the chemical potential and
$\hbar\omega_n=\pi T(2n+1)$ are fermionic Matsubara frequencies.
The electronic bands $E_{{\bf k},n}$
are obtained from the diagonalization
of Eq. (\ref{hamiltonian}),
\begin{equation}
E_{{\bf k}}^2=\frac{\gamma^2}{2}+\frac{\Delta^2}{4}+(\hbar v k)^2\pm
\sqrt{\frac{\gamma^4}{4}+(\hbar v k)^2(\gamma^2+\Delta^2)},
\end{equation}
and are labeled according to
Fig. \ref{f-bands}. 

We can also define the current operator
$j_\alpha=(1/N)\sum_{{\bf k},\sigma}
\Psi_{{\bf k},\sigma}^\dagger \hat{j}_{{\bf k},\alpha} \Psi_{{\bf k},\sigma}$,
where $\alpha=x,y$, $N$ is the total number of unit cells,
and where
\begin{eqnarray}
\label{curr1}
\hat{j}_{{\bf k},\alpha}
&=&-\frac{e}{\hbar}
\frac{d}{dk_\alpha}\hat{H_{\bf k}}.
\end{eqnarray}
In particular,
focusing on the current operator along the $y$-axis,
from Eqs. (\ref{h0})-(\ref{hamiltonian}) we get
\begin{eqnarray}
\label{curr}
\hat{j}_{{\bf k},y}
&=&
-e v \hat{I}(\hat{\sigma}_y),
\end{eqnarray}
where $\hat{A}(\hat{B})\equiv \hat{A}\otimes \hat{B}$.

The electron-phonon Hamiltonian describing electrons interacting
with the optical modes has also been discussed by several
authors.\cite{ando2L,Ando_09,gava}
As mentioned above,
in bilayer graphene there are two optical in-plane phonons at ${\bf q}=0$
(see Fig. \ref{f-basis}):
the $E_g$ symmetric (S) mode  and the antisymmetric (A)  $E_u$ one.
In the absence of a potential difference
between the two layers, the first one is Raman active while the
second one is infrared active.  The relative displacement of the two
sublattice atoms A and B in the first layer
is given for example by:
\begin{equation}
{\bf u}_\nu({\bf r})=
\sum_{{\bf q},\mu}
\sqrt{\frac{\hbar^2}{4M\hbar\omega_0}}
(c_{{\bf q},\mu}+c_{-{\bf q},\mu}^\dagger){\bf e}_\mu({\bf q})
\mbox{e}^{i{\bf q}\cdot {\bf r}}.
\end{equation}
$M$ is the mass of a
carbon atom, $\omega_0$ is the phonon frequency at the $\Gamma$ point,
$\mu=t,l$ denotes the polarization (transverse or longitudinal) and $
c_{{\bf q},\mu}$ and $c^\dagger_{{\bf q},\mu}$ are the phonon creation and
annihilation operators respectively. Using $q_x=q \cos \phi({\bf q})$ and
$q_y=q \sin \phi({\bf q})$ the polarization vectors ${\bf e}_\mu({\bf q})$
are given by ${\bf e}_l({\bf q})=i(\cos \phi({\bf q}),\sin \phi({\bf q}))$
and ${\bf e}_t({\bf q})=i(-\sin \phi({\bf q}), \cos \phi({\bf q}))$. 
Following Ref. \onlinecite{ando2L} the interaction between the optical
phonons and the electrons at the $K$ point can be written as 
\begin{equation}
H_{int}=-\sqrt{2}\frac{\beta \hbar v}{b^2} \sigma^{(\pm)}\times {\bf u}_\nu
({\bf r}),
\end{equation}
where $\sigma^{(+)}=\hat I(\hat \sigma)$, $\sigma^{(-)}=\hat \sigma_z (\hat
\sigma)$, $b=1.42$ \AA\, is the in-plane carbon-carbon nearest neighbor
distance, and $\beta$ is a dimensionless parameter related to the
deformation potential, whose typical value is $\beta=2.88$.\cite{piscanec,lazzeri}
In the following we shall consider the case of an electric field along the
$y$-axis, so that only the lattice vibrations along the $y$ direction will
couple to the light. Since for ${\bf q}\rightarrow 0$ the result is
independent on $\phi({\bf q})$, we take $\phi({\bf q})=0$, so that only the
${\bf e}_t$ polarization vector has a component along $y$. As a consequence
we can write, close to the K point,
the electron-phonon interaction for the $\nu=$ A,S mode as:
\begin{eqnarray}
\label{hint}
H_\nu
&=&
\sum_{{\bf k},\sigma} \Psi^\dagger_{{\bf k+q},\sigma}
\hat{V}_{\nu}({\bf q}) \Psi_{{\bf k},\sigma}e^{i{\bf q}\cdot {\bf r}}
(c_{{\bf q},\mu}+c_{-{\bf q},\mu}^\dagger),
\end{eqnarray}
where 
\begin{eqnarray}
\label{va}
\hat{V}_{A}({\bf q}\rightarrow 0)
&=&i g \hat \sigma_z(\hat \sigma_x),\\
\label{vs}
\hat{V}_{S}({\bf q}\rightarrow 0)
&=&i g \hat I(\hat \sigma_x),
\end{eqnarray}
and where 
$g=(\beta \hbar v/b^2)\sqrt{\hbar^2/2M\hbar\omega_0}=0.27$ eV.
Note that, since ${\bf e}_t({\bf q})=-{\bf e}_t(-{\bf q})$, one has that
$\lim_{{\bf q}\rightarrow 0}\hat V_\nu(-{\bf q})=- \lim_{{\bf q}\rightarrow
0}\hat V_\nu({\bf q})$.

We have now all the tools to investigate the full optical properties.
In this regard
it is convenient to define the generic propagator
for bosonic operators $A$ and $B$ as:
\begin{eqnarray}
\chi_{AB}(i\omega_m)
&=&
-\int_0^{1/T}d\tau \langle
T_\tau A(\tau)B(0)\rangle\exp[i\omega_m \tau],
\label{chiab}
\end{eqnarray}
where $\tau$ is the imaginary time, $T_\tau$ the time-ordering operator
and $\hbar\omega_m=2\pi m T$ the bosonic
Matsubara frequency.

The complex electronic optical conductivity per layer $\sigma(\omega)$
is obtained from the analytical continuation
of $\chi_{jj}(i\omega_m)$ to the real axis 
($i\omega_m \rightarrow \omega+i0^+$):
\begin{eqnarray}
\label{optc}
\sigma(\omega)
&=&
-\frac{\hbar}{V^{\rm 3D}} \frac{\chi_{jj}(\omega+i0^+)}{i\hbar \omega},
\end{eqnarray}
where $V^{\rm 3D}=2dS_{\rm cell}^{\rm 2D}$ with $d=3.35$ \AA\, being the
interlayer distance and  $S_{\rm cell}^{\rm 2D}=\sqrt{3}a^2/2$ the 
two-dimensional
area of the graphene unit cell, with $a=2.46$ \AA.

At the lowest non-interacting order, the current-current response function
$\chi_{jj}(i\omega_m)$  reduces to the single-bubble approximation
depicted in Fig. \ref{f-diagrams}a:
\begin{eqnarray}
\label{chi-irr}
\chi_{jj}(i\omega_m)
&=&
N_s N_v\frac{T}{N}\sum_{{\bf k},n}
\mbox{\rm Tr}
\Big[
\hat{j}_y
\hat{G}_0({\bf k},i\omega_n+i\omega_m)
\nonumber\\
&&
\times
\hat{j}_y
\hat{G}_0({\bf k},i\omega_n)
\Big],
\end{eqnarray}
where $N_s=N_v=2$ are the spin and valley degeneracies, respectively,
and $\hat{G}_0({\bf k},i\omega_n)$
is the $4 \times 4$ non-interacting electron Green's function which,
in the orbital basis, is not diagonal. As a consequence, 
the retarded response function $\chi_{jj}(\omega)$ will have contributions
coming from both the interband and intraband electronic excitations. In general
it can be decomposed as
\begin{eqnarray}
\chi_{jj}(\omega)
&=&\sum_{\alpha,\beta=1}^4\chi_{jj}^{\alpha\beta}(\omega),
\label{chimix}
\end{eqnarray}
where $\alpha,\beta$ are the band indexes and
$\chi_{jj}^{\alpha\beta}(\omega)$ describes the intraband ($\alpha=\beta$)
or interband ($\alpha\neq \beta$) particle-hole excitations contributing to
the total $\chi_{jj}$.  The properties of the optical conductivity in gated
bilayer graphene has been theoretically investigated in details in
Ref. \onlinecite{Nicol} and
experimentally confirmed in Refs. \onlinecite{Kuzmenko1,Mak,Zhang,Kuzmenko2}.
One can see that, at the phonon energies
$\omega \approx 0.2$ eV,
only the 2-3 interband transitions contribute to $\chi_{jj}''(\omega)$
(and therefore to the optical electronic background).\cite{Nicol}
Such background could be modulated by the charge
doping (and therefore by the applied gate voltage) so that it drastically
vanishes for $2|\mu| \gtrsim 0.2$ eV. 
Within the commonly widespread idea,
one would expect that this is the electronic background that controls the
$q$ Fano parameter and then the asymmetry of the phonon peak. 
However, as we shall see below, a correct application of Fano
theory to graphene leads to a different characteristic electronic response
function related to the Fano effect seen in several experiments.

Optical properties of the phonon resonances can be investigated
with the charged-phonon theory by analyzing
the diagrams in  Fig. \ref{f-diagrams}b.
In ungated samples, only one phonon, the $E_u$ antisymmetric mode,
is expected to be IR active.
We can write thus
\begin{eqnarray}
\Delta\chi_{jj}(\omega)
&=&
\chi_{j{\rm A}}(\omega)D_{\rm AA}(\omega)
\chi_{{\rm A}^\dagger j}(\omega),
\label{chiriceAA}
\end{eqnarray}
where $\chi_{j{\rm A}}$ is the mixed response function
between the current operator ${j}_y$
and the electron-phonon scattering operator ${V}_{\rm A}$.

However,  in the most general
case of gated samples with $\Delta \neq 0$,
also the symmetric $E_g$ mode acquires a finite IR activity.\cite{tang,noi}
We can write then
\begin{eqnarray}
\Delta\chi_{jj}(\omega)
&=&
\chi_{j{\rm A}}(\omega)D_{\rm AA}(\omega)
\chi_{{\rm A}^\dagger j}(\omega)
\nonumber\\
&&
+\chi_{j{\rm S}}(\omega)D_{\rm SS}(\omega)
\chi_{{\rm S}^\dagger j}(\omega)
\nonumber\\
&&
+\left[
\chi_{j{\rm A}}(\omega)D_{\rm AS}(\omega)
\chi_{{\rm S}^\dagger j}(\omega)
+\mbox{h.c.}
\right],
\label{chirice}
\end{eqnarray}
where $\nu=$A, S.

Eqs. (\ref{chirice}) can be
completed with the Dyson's equation for the phonon Green's functions
$D_{\nu\nu'}$:\cite{Ando_09,gava}
\begin{eqnarray}
\label{phdyson}
\left[D^{-1}(\omega)\right]_{\nu\nu'}
&=&
\delta_{\nu,\nu'}\left[D^{-1}_0(\omega)\right]
-\chi_{\nu^\dagger \nu'}(\omega),
\end{eqnarray}
where $D_0(\omega)=
-2\hbar\omega_0/[\hbar^2\omega_0^2-(\hbar\omega+i0^+)^2]$ is the
phonon propagator in the absence of electron-phonon interaction and
$\chi_{\nu^\dagger \nu'}(\omega)$
provide the matrix components of the
phonon self-energy. 
The quantity
$\omega_0$ represents the bare phonon frequencies for a generic mode $\nu$
in the absence of electron-phonon interaction, and it is
assumed here to be degenerate $\omega_{0{\rm A}}=\omega_{0{\rm S}}=\omega_0$.
As it has been discussed in detail in Refs. \onlinecite{Ando_09,gava,yan3},
the mixing between the A and S modes (mediated by the self energy
$\chi_{\rm AS}$) is only active when $\Delta \neq 0$. In this case the
phonon eigenmodes do not correspond any more to symmetric/antisymmetric
vibrations of the atoms in neighboring layers, so that each A and S
propagator has a double-pole structure, centered at the values $\omega_\pm$
of the phonon eigenfrequencies.

Eqs. (\ref{chirice})-(\ref{phdyson})
provide the theoretical tools needed
to evaluate microscopically the onset and the properties
of phonon peaks in the optical conductivity. 
In general, all the information about {\em frequencies} and {\em lifetimes} of phonons
is encoded in the phonon self-energy $\chi_{\nu^\dagger \nu'}$
whereas the mixed (current/electron-phonon interaction) response
functions $\chi_{j\nu}$ are related to the {\em intensity}
and to the possible {\em Fano asymmetry} of the phonon peaks, as we shall discuss
in the next Section.

\section{Correspondence with the standard Fano theory}
\label{s:fano}

In order to better clarify the connection between Eq. (\ref{chirice}) and 
the standard Fano theory let us consider first the case 
$\Delta=0$, where only the A mode is optically active
and the expression (\ref{chirice}) reduces to
Eq. (\ref{chiriceAA}).
In addition, for $\Delta=0$ the mixed phonon self-energy
$\chi_{{\rm A}^\dagger {\rm S}}$ vanishes, which means that the A
and S modes coincide with the eigenmodes for the lattice vibrations. In
particular, this implies that the phonon propagator
$D_{\rm AA}$ has a single resonance, and the 
main effect of the phonon self-energy is to induce 
a shift of the phonon frequency, $\hbar\omega_{\rm A}=
\hbar\omega_0+\mbox{Re}\chi_{\rm A^\dagger A}(\omega_{\rm A})$
and a finite line broadening $\Gamma_{\rm A}=
-\mbox{Im}\chi_{\rm A^\dagger A}(\omega_{\rm A})$.
Therefore, for $\omega \approx \omega_{\rm A}$ we can approximate
the phonon propagator as
\begin{equation}
\label{daa}
D_{\rm AA}(\omega)=\frac{1}{
\hbar(\omega-\omega_{\rm A})+i\Gamma_{\rm A}}.
\end{equation}
Using the relation
$\chi_{{\rm A}^\dagger j}(\omega)=\chi_{j{\rm A}}(\omega)$, the 
optical conductivity can be expressed in terms of the
real and imaginary part of the mixed response function $\chi_{jA}$ and of
the propagator (\ref{daa}). In particular, 
introducing the variable 
$z=\hbar[\omega-\omega_{\rm A}]/\Gamma_{\rm A}$ we have
Re$D_{\rm AA}=
D_{\rm AA}'=z/\Gamma_{\rm A}(1+z^2)$ and 
Im$D_{\rm AA}=D_{\rm AA}''=-1/\Gamma_{\rm A}(1+z^2)$, so that
\begin{eqnarray}
{\rm Im} \Delta
\chi_{jj}&=&D_{AA}''[(\chi_{jA}')^2-(\chi_{jA}'')^2]
+2D_{AA}'\chi_{jA}'\chi_{jA}''
\nonumber
\\
&=&
-\frac{
(\chi_{jA}')^2-(\chi_{jA}'')^2-2z  \chi_{jA}'\chi_{jA}''
}{\Gamma_{\rm A}(1+z^2)}.
\end{eqnarray}
As a consequence, the real part of the optical conductivity (\ref{optc}) 
close to the resonance frequency $\omega_{\rm A}$ can be written as:\cite{cbk}
\begin{eqnarray}
\label{fanorice1}
\mbox{Re}\Delta\sigma(\omega)\Big|_{\omega\approx \omega_{\rm A}}
&\approx&
I_{\rm A}
\left[
\frac{q_{\rm A}^2-1+2q_{\rm A} z}{q_{\rm A}^2(1+z^2)}
\right],
\end{eqnarray}
where we defined the Fano parameter $q$ as
\begin{eqnarray}
\label{fanorice2}
q_{\rm A}
&=&
-\frac{\chi_{j{\rm A}}'(\omega_{\rm A})}
{\chi_{j{\rm A}}''(\omega_{\rm A})},
\end{eqnarray}
while the prefactor is given by
\begin{eqnarray}
\label{ia}
I_{\rm A}
&=&
\frac{ \left[\chi_{j{\rm A}}'(\omega_{\rm A})\right]^2}
{\omega_{\rm A}\Gamma_{\rm A}V}.
\end{eqnarray}
As one can see, Eq.\ (\ref{fanorice1}) reproduces the Fano formula, where
the $q$ parameter controls the asymmetry of the peak with respect to a
standard Lorentzian profile, that is recovered in the limit of
$q\rightarrow \infty$. As observed already in Ref. \onlinecite{cbk},
the derivation of Eqs. (\ref{fanorice1})-(\ref{fanorice2}) shows that the
Fano effect stems from a correct implementation of the
charged-phonon theory.
Therefore, for the sake of simplicity, in the following we shall
regards the phonon properties (intensity and lineshape asymmetry)
arising from this common nature as
 the ``Fano-Rice'' effect.
Moreover, the above set of equations provides a general scheme
to calculate microscopically the relevant parameters that control
the shape and the intensity of the phonon peak, in particular $q_{\rm A}$
and $I_{\rm A}$, that are fully determined once that the mixed response
function $\chi_{j{\rm A}}$ is computed.

Before showing explicitly the calculation of $\chi_{jA}$ we would like to
make a more direct comparison with the standard
Fano formalism\cite{fano}
that is often quoted in the literature. Following the original work
by Fano,\cite{fano} the asymmetry parameter $q$ that measures the interference
effect between a discrete phonon state $|\phi\rangle$ of energy 
$\omega_0$ and a continuum of electronic states $|\psi_\omega\rangle$
can be written as
\begin{eqnarray}
\label{defq}
q
&=&
\frac{\langle \phi | T | i \rangle+P\int d \omega' 
\frac{\displaystyle V_{\omega'}\langle \psi_{\omega'} | T | i \rangle}
{\displaystyle \omega_0-\omega'}}
{\langle \psi_{\omega_0} | T | i \rangle \pi V_{\omega_0}},
\end{eqnarray}
where 
$P$ denotes the principal part of the integral, and
$V_\omega$ measures the hybridization between the phonon and the
electronic states at the energy $\omega$,
$V_\omega= \langle \psi_\omega | H | \phi \rangle$. Here
$\langle f|T|i\rangle$ denotes in general the transition amplitude from an
initial state $|i\rangle$ and a final state $|f\rangle$. The
first term in the numerator of Eq.\ (\ref{defq}) represents the response of
the bare phonon state, {\em i.e.} the bare dipole of the system under the lattice
distortion, while the second one gives the contribution coming from the
electronic excitations. Note that the relevant electronic excitations
are not restricted to the vicinity of the phonon frequency $\omega_0$ but 
they involve higher energy states $\omega'$ as well.
In contrast, the denominator depends solely on
the processes at $\omega=\omega_0$ and it
vanishes if there is no
electronic continuum at the phonon energy, so that
in this case $q=\infty$ and no Fano
asymmetry is expected. In ordinary systems 
the bare phonon intensity $\langle \phi | T | i \rangle$ is large,
so that $q$ becomes appreciably small only in the presence of a
considerable electron-phonon coupling $V_{\omega_0}$, and the observation of
a pronounced Fano asymmetry is considered as a signature of large
electron-phonon interactions.\cite{fanocardona}
However, the case of graphene is radically
different: since here the bare
phonon activity is negligible, the main phonon intensity comes from
the particle-hole excitations (Rice effect), and the Fano asymmetry can be
pronounced even in the presence of a relatively small electron-phonon
coupling. 
In order to show this more explicitly, 
we introduce the complex function:
\begin{eqnarray}
\chi(\omega)
&=&
\label{defchi}
\int d\omega' \frac{ C({\omega}')}{\omega-\omega'+i\eta}, 
\end{eqnarray}
where we defined
\begin{equation}
\label{defc}
C(\omega)=V_{\omega}
\langle \psi_\omega | T | i \rangle .
\end{equation}
For $\langle \phi | T | i \rangle\approx 0$, Eq.\ (\ref{defq})
can be written as:
\begin{equation}
\label{noiq}
q=-\frac{\chi'(\omega_0)}{\chi''(\omega_0)},
\end{equation}
which reduces to Eq.\ (\ref{fanorice2}) derived above from the
charged-phonon
theory. Note that both $\chi'$ and $\chi''$ are proportional to the
electron-phonon matrix element $V_\omega$ ({\em i.e.} $g$ in our previous notation), so that
the strength of the electron-phonon interaction
cancels out in Eq. (\ref{noiq}) as well as in (\ref{fanorice2}).
As we shall see below, the response function $\chi_{j{\rm A}}$ has
exactly the form of Eq.\ (\ref{defchi}), and we will be able
to compute explicitly the
function $C(\omega)$.

\begin{figure}[t]
\begin{center}
\includegraphics[scale=0.40,clip=]{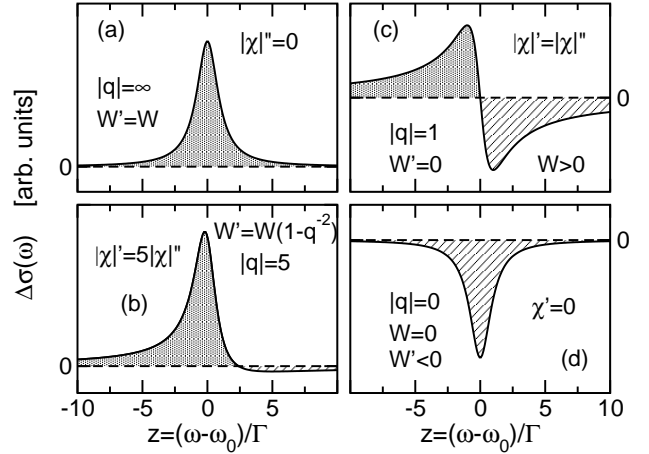}
\end{center}
\caption{Sketch of the optical properties of a phonon resonance
in different characteristic regimes: (a)
positive symmetric Lorentzian peak
($|q|=\infty$; $|\chi"_{j\nu}(\omega_\nu)|=0$),
where $W=W'$;
(b) weak asymmetric Fano profile
(ex.: $|q|=5$; $|\chi'_{j\nu}(\omega_\nu)|=5|\chi"_{j\nu}(\omega_\nu)|$);
 (c) highly asymmetric case ($|q|=1$;
$|\chi'_{j\nu}(\omega_\nu)|=|\chi"_{j\nu}(\omega_\nu)|$),
where
the integrated area $W'=0$ because of the cancellation
of positive and negative spectral regions;
(d)
negative phonon peak ($|q|=0$;
$\chi'\approx 0$), where
the ``bare'' intensity vanishes, $W=0$,
although a phonon anomaly is visible in the imaginary
part of $\chi$. The total phonon strength $P$ is the same in all the cases.
}
\label{f-fano}
\end{figure}

Few more final observations are in order concerning Eqs.\
(\ref{fanorice1})-(\ref{wa}).
First, in contrast to the ordinary case where
also the bare dipole charge of the phonon must be considered, in the case
of graphene the two quantities $q_{\rm A}$, $I_{\rm A}$ are not
independent. This is again a consequence of a correct implementation
of the charged-phonon effect, i.e. of the
fact that in graphene the optical activity of the phonon is fully borrowed
from the electronic excitations. 
This also permits to quantify the ``strength'' of a phonon
resonance in the optical conductivity on a more rigorous ground.
To this aim, for a given phonon mode $\nu$,
two typical quantities are considered in literature:
the integrated spectral area
\begin{eqnarray}
\label{wap}
W'_\nu=\int d\omega \mbox{Re}\Delta\sigma(\omega),
\end{eqnarray}
and the phonon intensity $W_\nu$, as obtained
from the Fano-like fit in Eq. (\ref{fanorice1})
as
\begin{equation}
\label{wa}
 W_\nu=\pi \Gamma_\nu I_\nu.
\end{equation}
The quantity $W_\nu$ is considered to be the ``bare''
spectral intensity that the mode would have in the absence
of the Fano interference.
It is straightforward to show that these two quantities are
related by the formula
\begin{eqnarray}
\label{wapp}
W_\nu'
=
\left(1-\frac{1}{q_\nu^2}\right)W_\nu,
\end{eqnarray}
so that $W' \rightarrow W_\nu$ when $|q_\nu| \rightarrow \infty$
for a symmetric Lorentzian profile (Fig. \ref{f-fano}a).

Both $W_\nu$ and  $W_\nu'$ can be expressed in the common
Fano-charged-phonon framework via the mixed response function $\chi_{j\nu}$:
\begin{eqnarray}
\label{wap2}
W_\nu
=
\frac{\pi \chi'^2_{j\nu}(\omega_\nu)}{\omega_\nu},
\end{eqnarray}
and
\begin{eqnarray}
\label{wap22}
W_\nu'
=
\frac{\pi \left[\chi'^2_{j\nu}(\omega_\nu)-\chi''^2_{j\nu}(\omega_\nu)
\right]}{\omega_\nu}.
\end{eqnarray}

It should be however stressed that none of these two quantities,
$W$, $W'$, can provide a satisfactory quantification
of the {\em magnitude} of the phonon optical anomaly.
For instance,
in the case $|q_\nu| \approx 1$,
where the phonon peak asymmetry is strongest and
which corresponds to
$|\chi'_{j\nu}(\omega_\nu)|=|\chi''_{j\nu}(\omega_\nu)|$,
we get $W_\nu'=0$ due to the cancellation of positive and negative
spectral regions [Fig. \ref{f-fano}c].
On the other hand, when $\chi_{j\nu}'(\omega_\nu)\approx 0$,
the spectral properties are characterized by
a sizable {\em negative} Lorentzian peak
whose intensity is driven by $\chi_{j\nu}''(\omega_\nu)$
whereas the estimated ``bare'' intensity is vanishing small,
$W_\nu\approx 0$ [Fig. \ref{f-fano}d].

Both definitions $W$ and $W'$ thus fail to describe
the actual magnitude of the optical phonon resonance
independently of its Fano-like properties.
This problem can be solved however thanks
to the microscopical identification
of the optical properties in terms of the mixed
response function $\chi_{j\nu}$.
The simple identification of $W$ and $W'$
as $W_\nu\propto \chi'^2_{j\nu}$,
$W_\nu'\propto \chi'^2_{j\nu}-\chi''^2_{j\nu}$
suggests us to introduce a strictly {\em positively defined} quantity,
\begin{eqnarray}
\label{pstrength}
P_\nu
=
\frac{\pi \left[\chi'^2_{j\nu}(\omega_\nu)+\chi''^2_{j\nu}(\omega_\nu)
\right]}{\omega_\nu},
\end{eqnarray}
which we refer to as {\em phonon strength}.
Note that $P \rightarrow 0$ only when both
$\chi'_{j\nu}, \chi''_{j\nu}\rightarrow 0$, {\em i.e.}
when the phonon features are indeed vanishingly small.
In the following, when discussing the {\em magnitude}
of a optical phonon resonance, we shall therefore refer
to this quantity $P_\nu$, which permits to
describe the actual visibility of a phonon structure
independently of its Fano shape.
The robustness of the parameter $P_\nu$ to characterize
the magnitude of the phonon resonance
independently of its lineshape and its Fano properties
is demnostrated in Fig. \ref{f-fano} where all
the panel where evaluated for fixed phonon strength.

As a last point we would like to stress that
it is the mixed response function $\chi_{j{\rm A}}$,
and not the electronic background $\chi_{jj}$,
as it was considered in Refs. \onlinecite{rice,rice2},
that
determines the phonon strength and the asymmetry.
As we shall see this
makes a crucial difference. Indeed, while all the possible electronic
excitations contribute to $\chi_{jj}$, only a subset of them enters into the
mixed response function $\chi_{j{\rm A}}$ (and $\chi_{j{\rm S}}$ in
the case $\Delta \neq 0$), determining in this way the exact selection
rules for the phonon activity in bilayer graphene.

\section{Fano-Rice properties in ungapped 
bilayer graphene ($\Delta=0$)}
\label{s-nogap}

In the case of unbiased graphene the mixed response function 
$\chi_{j{\rm A}}$ can be computed analytically in the
bare-bubble approximation, namely:
\begin{eqnarray}
\label{chi-jn}
\chi_{j\nu}(i\omega_m)
&=&
N_s N_v\frac{T}{N}\sum_{{\bf k},n}
\mbox{\rm Tr}
\Big[
\hat{j}_y
\hat{G}({\bf k},i\omega_n+i\omega_m)
\nonumber\\
&&\times
\hat{V}_\nu
\hat{G}({\bf k},i\omega_n)
\Big],
\end{eqnarray}
where we use in Eq.\ (\ref{chi-jn})
the bare electronic Green's functions.
Some additional
details of the calculation are given
in Appendix \ref{app-chijA}. 
Due to the 
multiband structure of the system,
$\chi_{j{\rm A}}$ has the typical structure of a particle-hole Lindhard response
function, with proper coherence factors $C^{nm}_{j{\rm A}}$ weighting the
contributions of the various excitations between the $n$ and
$m$ bands. In particular, using the explicit
matrix expressions of the $\hat j$ and $\hat V_{\rm A}$ operators, one gets
\begin{eqnarray}
\label{chijAmix}
\chi_{j{\rm A}}(\omega)
&=&
\chi_{j{\rm A}}^{12}(\omega)+\chi_{j{\rm A}}^{13}(\omega)
-\chi_{j{\rm A}}^{24}(\omega)-\chi_{j{\rm A}}^{34}(\omega).
\end{eqnarray}
Here $n,m$ are the band indexes and
$\chi^{nm}_{jA}(\omega)=\pi^{nm}_{jA}(\omega)-
\pi^{mn}_{jA}(\omega)$, where
\begin{eqnarray}
\label{elements}
\pi_{j{\rm A}}^{nm}(\omega)
&=&
\frac{1}{N}
\sum_{\bf k}
 C_{j{\rm A},{\bf k}}^{nm}
\frac{f(E_{{\bf k},n}-\mu)-f(E_{{\bf k},m}-\mu)}
{E_{{\bf k},n}-E_{{\bf k},m}+\hbar\omega+i\eta},
\end{eqnarray}
and where
\begin{equation}
 C_{j{\rm A},{\bf k}}^{nm}=ge vN_s N_v
\frac{\gamma_1}{4\sqrt{(\hbar v k)^2+\gamma_1^2}}
\label{cfunc}
\end{equation}
for $(n,m)=(1,2), (1,3), (2,4), (3.4)$ and zero otherwise.
In Eq. (\ref{elements})
$\mu$ represents the chemical potential,
and $\eta$ is a phenomenological parameter
that accounts for the damping in the electronic states,
so that the clean limit corresponds to
$\eta \rightarrow 0$.\cite{noteeta}
Note that, once we identify
\begin{eqnarray}
C(\omega)
&=&
\frac{1}{N}
\sum_{m,n}\sum_{\bf k}
 C_{j{\rm A},{\bf k}}^{nm}\delta
(E_{{\bf k},n}-E_{{\bf k},m}+\hbar\omega)
\nonumber\\
& & \times 
[f(E_{{\bf k},n}-\mu)-f(E_{{\bf k},m}-\mu)],
\end{eqnarray}
the function $\chi_{jA}(\omega)$ can be written exactly as in Eq.\
(\ref{defchi}) above, with the particle-hole energy difference
$\omega'=E_{{\bf k},n}-E_{{\bf k},m}$ playing the role
of the electronic continuum in the Fano theory.  Thus, our approach allows
one to identify the optical intensity of the phonon peaks
and their Fano parameters in terms of the real and
imaginary parts of a specific response function, that can be calculated
using the standard diagrammatic theory.  This means in particular that we
can determine: ($i$) which phonon is coupled to the current, and ($ii$)
which electronic excitations couple to each phonon mode.

All these elements can be quantified in an analytical way
for the A
phonon in the case of no gap. First of all, note that for $\mu=0$ the term
$\chi_{j{\rm A}}^{13}$ cancels precisely with $\chi_{j{\rm A}}^{24}$,
whereas $\chi_{j{\rm A}}^{12}$, $\chi_{j{\rm A}}^{34}$ are both vanishing
because particle-hole excitations between completely full or empty
bands are not allowed. As a consequence $\chi_{j{\rm A}}(\omega)=0$,
implying that the
intensity of the A mode, although it has the correct symmetry, vanishes in
undoped bilayer graphene.\cite{noteTB}
As we shall see, this property holds true
even in the presence of an interlayer difference potential $\Delta$.

\begin{figure}[t]
\begin{center}
\includegraphics[scale=0.33,clip=]{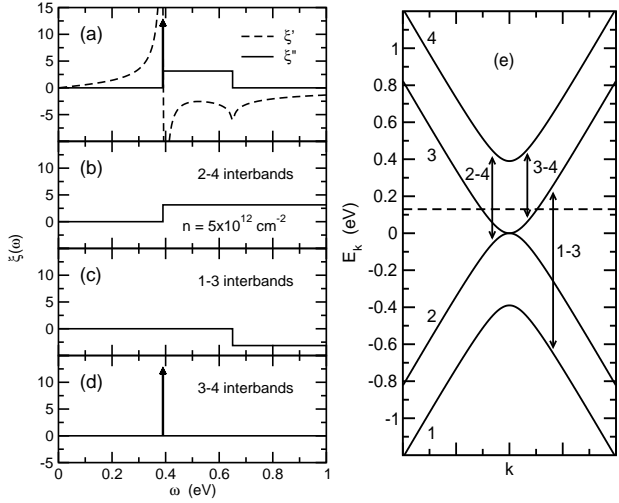}
\end{center}
\caption{(a) Real  (dashed line) and imaginary (solid line)
part of the  dimensionless response functions
$\xi(\omega)$ for $n=5 \times 10^{12}$ cm$^{-2}$ and $T=0$.
The vertical arrow for $\xi''(\omega)$
represents a $\delta$-function with spectral weight $2|\mu|$, see 
Eq. (\ref{imchi}). Panels (b)-(d): contributions of the single 
interband transitions to the total $\xi''(\omega)$.
Panel (e): sketch of the allowed particle-hole interband excitations.
The horizontal dashed line marks here the chemical potential.
}
\label{f-chijA}
\end{figure}

The total mixed response function $\chi_{j{\rm A}}(\omega)$
as well as its single contributions $\chi_{j{\rm A}}^{nm}(\omega)$
can be easily evaluated in the clean limit, $\eta=0$, and in the linear
approximation for the band dispersion,
corresponding to the Hamiltonian given by
Eq. (\ref{hamiltonian}). 
Details of the computation are also reported in
Appendix \ref{app-chijA}.
It is convenient to introduce the dimensionless
quantity $\xi(\omega)$:
\[
\chi_{j{\rm A}}(\omega)=A\xi(\omega),
\]
where $A=ge\gamma S_{\rm cell}^{\rm 2D}/4\pi\hbar^2 v$.
At $T=0$
we obtain the analytical expressions:
\begin{eqnarray}
\xi'(\omega)
&=&
\ln\left[\frac{(\gamma_1+\hbar\omega)(\gamma_1-\hbar\omega+2|\mu|)}
{(\gamma_1-\hbar\omega)(\gamma_1+\hbar\omega+2|\mu|)}\right]
\nonumber\\
&&
-\theta(|\mu|-\gamma_1)
\ln\left[\frac{(\gamma_1+\hbar\omega)(\gamma_1+\hbar\omega-2|\mu|)}
{(\gamma_1-\hbar\omega)(\gamma_1-\hbar\omega-2|\mu|)}\right]
\nonumber\\
\label{rechi}
&&
+\frac{4\hbar\omega 
[|\mu|\theta(\gamma_1-|\mu|)+\theta(|\mu|-\gamma_1)]}
{(\gamma_1^2-\hbar^2\omega^2)},
\\
\xi''(\omega)
&=&
\pi
\left\{
\theta(\hbar|\omega|-\gamma_1)\theta(\hbar|\omega|-2|\mu|+\gamma_1)
\right.
\nonumber\\
&&
-\theta(\hbar|\omega|-2|\mu|-\gamma_1)
\nonumber\\
\label{imchi}
&&\left.
+2
[|\mu|\theta(\gamma_1-|\mu|)+\theta(|\mu|-\gamma_1)]
\delta(\hbar|\omega|-\gamma_1)
\right\}.
\end{eqnarray}
The real and imaginary part of $\chi_{j{\rm A}}(\omega)$
for a representative case
$n=5 \times 10^{12}$ cm$^{-2}$ ($\mu=0.13$ eV)
are shown in Fig. \ref{f-chijA}a, 
whereas in panels (b)-(d) the contribution of the single 
interband transitions is also shown.
It is interesting to remark that for $|\omega| \gtrsim \gamma_1+2|\mu|$ the
term Im$\chi_{j{\rm A}}^{13}(\omega)$ cancels exactly with Im$\chi_{j{\rm
A}}^{24}(\omega)$, so that only a limited low-frequency energy-window
$\gamma_1 \lesssim |\omega| \lesssim \gamma_1+2|\mu|$ contributes to 
the imaginary part (and hence, via Kramers-Kronig relations, to the real part)
of the mixed response function $\chi_{j{\rm A}}$.  This fact guarantees
that the result is valid even if the calculation has been done using the
linearized graphene bands instead of the full periodic band structure.

According to Eq.\ (\ref{fanorice1}), the computation of
the optical phonon spectra requires
the evaluation of the quantities (\ref{rechi})-(\ref{imchi}) at the
phonon frequency $\omega_A$.
In order to determine $\omega_{\rm A}$, one should in principle
solve the phonon Dyson's equation (\ref{phdyson}).
However, the shift of the renormalized phonon
frequencies $\omega_\nu$ is only a few meVs, which is much smaller
than $\omega_0 \approx 200$ meV, so that we can
replace $\omega_{\rm A}$ with $\omega_0$.
In any case, as one can see in Eq.\ (\ref{chijAmix}), since
the low-energy 2-3 interband transitions are not allowed in $\chi_{j{\rm A}}$,
the lowest threshold for the particle-hole excitations is determined
in the clean limit $\eta=0$
by the 1-3 interband transitions with $\omega >\gamma_1=0.39$ eV, so that
Im$\chi_{j{\rm A}}(\omega_{\rm A})=0$.  According
to Eq.\ (\ref{fanorice2}) this implies that the A phonon peak in the clean
limit has no Fano asymmetry ($|q_{\rm A}|=\infty$), and the bare phonon
peak intensity $W_{\rm A}$ (\ref{wa}) coincides with the integrated area
($W_{\rm A}=W_{\rm A}'$) [see Eq.(\ref{wap})]. From Eq.\ (\ref{wa})
we get :
\begin{equation}
W_{\rm A}
=
\frac{\pi A^2}{\omega_{\rm A} V^{\rm 3D}}
\xi'^2(\omega_{\rm A})
=
{\lambda}\sigma_0\frac{\gamma_1^2}{\hbar \omega_{\rm A}}
\xi'^2(\omega_{\rm A}),
\end{equation}
where $\sigma_0=e^2/4\hbar d \approx 1816$ $\Omega^{-1}$ cm$^{-1}$ 
and $\lambda=(\sqrt{3}/\pi)g^2/(\hbar v/a)^2$ is the dimensionless
phonon coupling which, using the value $g=0.27$ eV estimated in Section \ref{s:rice},
results in $\lambda= 6\times10^{-3}$,\cite{ando2L} in
agreement with the experimental estimates given in Refs. \onlinecite{malard,noi}.
Finally, we obtain that $\pi A^2/\omega_{\rm A} V^{\rm 3D}=4.35\times 10^3$
$\Omega^{-1}$cm$^{-2}$.

\begin{figure}[t]
\begin{center}
\includegraphics[scale=0.35,clip=]{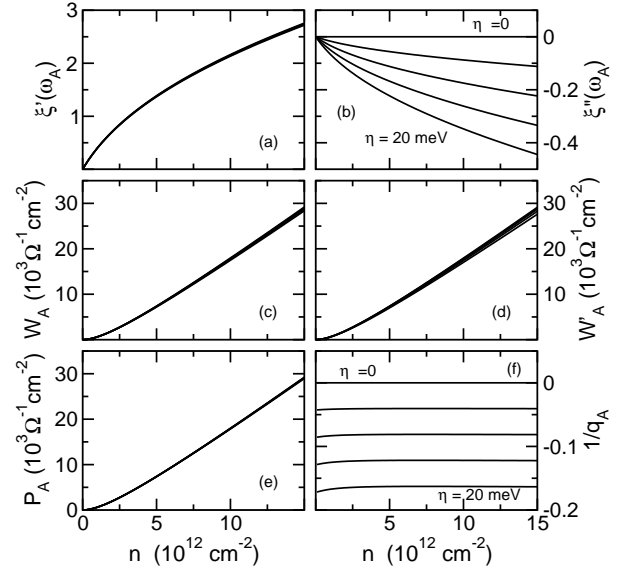}
\end{center}
\caption{Dimensionless quantities
$\xi'(\omega_{\rm A})$, $\xi''(\omega_{\rm A})$,
as well as the spectral weights $W_{\rm A}$, $W_{\rm A}'$,
$P_{\rm A}$ and the Fano factor $q_{\rm A}$
as functions of the charge concentration $n$
for the clean limit ($\eta=0$) and for finite damping
$\eta= 5, 10, 15, 20$ meV.
Curves for $\xi'(\omega_{\rm A})$, $W_{\rm A}$, $W_{\rm A}'$,
$P_{\rm A}$ at zero and finite $\eta$ are barely distinguishable.}
\label{f-wawa_n}
\end{figure}

To elucidate the doping dependence of the optical properties
of the A mode, we
show in  Fig.~\ref{f-wawa_n} the quantities
$\xi'(\omega_{\rm A})$, $\xi''(\omega_{\rm A})$
as functions 
of the charge doping $n$, along with the parameters
$W_{\rm A}$, $W'_{\rm A}$, $P_{\rm A}$ and $q_{\rm A}$.
As mentioned above, due to the absence of 2-3 interband transitions,
the imaginary part $\xi(\omega)$ is zero at $\omega=\omega_{\rm A}$
in the clean limit $\eta=0$ and relatively small for finite
$\eta$ (Fig. \ref{f-wawa_n}b),
so that the spectral weight is dominated by the real part
$\xi(\omega_{\rm A})$.
In this context the small contribution
at $\omega=\omega_{\rm A}$ for finite $\eta$
coming from the gapped 2-4 and 3-4 interband excitations
gives rise to a finite Fano factor $|q_{\rm A}|< \infty$ 
(Fig. \ref{f-wawa_n}f)
but it does not affect sensibly the spectral weights
$W_{\rm A}$, $W'_{\rm A}$ and the spectral strength $P_{\rm A}$.
In particular, the doping dependence of these latter quantities,
in the range $n$ here considered, is dominated by the
3-4 interband transitions [the third term in Eq. (\ref{rechi})],
where $\xi'(\omega_{\rm A})\propto |\mu|$,
so that, for $\xi''(\omega_{\rm A})\ll \xi'(\omega_{\rm A})$,
$W_{\rm A}=W'_{\rm A}=P_{\rm A} \propto [\xi'(\omega_{\rm A})]^2 \propto |\mu|^2 \approx n$.

Note also that, although the imaginary part $\xi''(\omega_{\rm A})$
does not contribute to the spectral weights 
$W_{\rm A}$, $W'_{\rm A}$ and $P_{\rm A}$,
it determines the magnitude of the
Fano asymmetry factor,
as shown in Fig. \ref{f-wawa_n}f.
As mentioned above, this is triggered by the finite spectral weight
in $\xi''(\omega_{\rm A})$ due to the broadening $\eta$ of the
higher-energy transitions. Increasing
the charge concentration $n$,
and hence the chemical potential $|\mu|$,
leads to an overall increase of
$\xi''(\omega_{\rm A})$ and, since $\xi'(\omega_{\rm A})$ 
is related by the Kramers-Kronig relations to the
low energy part of $\xi''(\omega)$, to a similar increase
of $\xi'(\omega_{\rm A})$. The charge-doping dependence 
of the magnitude of $\xi'(\omega_{\rm A})$ and $\xi''(\omega_{\rm A})$
is thus similar, making the Fano factor $q_{\rm A}$
almost independent
of $n$.
As we are going to see in the next Section,
the situation is different for the symmetric S mode,
when its infrared activity is triggered by a finite energy difference $\Delta \neq 0$.

\section{Fano-Rice properties in gapped ($\Delta\neq 0$)
bilayer graphene}
\label{s-gap}

In the above Section we have addressed the optical properties
of the antisymmetric $E_u$ mode in bilayer graphene in the absence
of any electrostatic potential gradient $\Delta$
between the two layers.
However, in most cases, the gating of the samples as well
as the influence of the
substrate give rise to a finite
potential difference between the two layers. In this case, the
antisymmetric A and symmetric S modes depicted in Fig. \ref{f-basis}
are no more eigenvectors of the lattice dynamics although they still represent
a suitable basis to investigate the optical properties
of bilayer graphene.
According to Eq. (\ref{chirice})
we can distinguish three different channels responsible
for the onset of phonon peaks in the infrared conductivity:
the direct coupling of the electron current with the A mode,
which is already present for $\Delta=0$
[first term of Eq. (\ref{chirice})];
the direct coupling with the S mode, which is induced by the
presence of the gap $\Delta$ and vanishes for $\Delta \rightarrow 0$
[second term of Eq. (\ref{chirice})];
the mixed A-S mode optical coupling where the incoming light
first excites the A lattice vibrations, which develop a
S component due to the hybridized A-S phonon self-energy, and
finally the light is re-emitted through the coupling of the A mode to the current
[third term of Eq. (\ref{chirice})].

\begin{figure}[t]
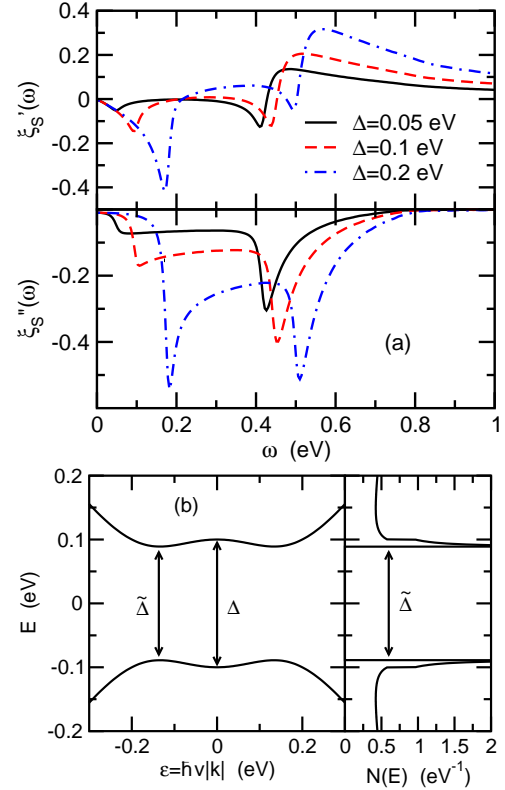

\begin{center}
\includegraphics[scale=0.49,clip=]{chijS-2.eps}
\hspace{3mm}
\includegraphics[scale=0.45,clip=]{f-bandgap.eps}
\end{center}
\caption{(a) Real part (upper panel) and imaginary
part (lower panel) of the  dimensionless response functions
$\xi(\omega)$ for $\mu=0$ and
for different values of the gap:
$\Delta=0.05, 0.1, 0.2$ eV.
Here we set $T=10$ K and $\eta=10$ meV.
(b) Gapped dispersion (left panel)
and corresponding density of states $N(\epsilon)$
(right panel) for $\Delta=0.2$ eV.
The vertical arrows mark the gap $\Delta$ at ${\bf k}=0$ (K point)
and the optical gap $\tilde{\Delta}$
at the bottom of the mexican hat. In this case the DOS
presents a divergent behavior.
}
\label{f-chijS}
\end{figure}

\subsection{Undoped case $n=0$, $\Delta \neq 0$}

An interesting case to elucidate the role
of the gap in triggering on the S-channel is
the undoped case ($\mu=0$) where the
antisymmetric mode $E_u$ is not involved
[$\chi_{j{\rm A}}^{\rm  irr}(\omega)=0$],\cite{noteTB}
and the only coupled lattice mode is the symmetric $E_g$ one
induced by the finite $\Delta$.
Such condition has been experimentally realized in Ref. \onlinecite{tang}.
In Fig. \ref{f-chijS}a we show, for different values of $\Delta$,
the real and imaginary part
of the dimensionless quantity $\xi_{\rm S}(\omega)=\chi_{j{\rm S}}(\omega)/A$, which
rules the optical properties of the phonon resonance.
An important difference here is that  the low-energy interband
transitions 2-3, that were forbidden in $\chi_{j{\rm A}}$,
are allowed in the response function $\chi_{j{\rm  S}}$.
The imaginary part of $\xi_{\rm S}(\omega)$ is thus finite
at low energies (limited only by the opening of the optical gap $\tilde\Delta$)
providing a finite Fano interference as long as $\omega_{\rm S} \le
\tilde{\Delta}$.
The sharp features in $\xi_{\rm S}''(\omega)$ are reflected
in peaked structures in the real part $\xi_{\rm S}'(\omega)$.
In particular, the structure at $\omega=0.4-0.5$ eV
can be associated with the transitions between bands
2 and 4, 
with lowest characteristic energy at $\approx \gamma_1+\tilde\Delta/2$, 
whereas the structure at low
energies
$\omega=0.05-0.2$ eV reflects the opening of the band gap on the
transitions 2-3. 
It is worth to note that, due to the mexican-hat shape of
the electronic dispersion, the lowest energy threshold is not
determined by the gap $\Delta$ at the K point but
by the actual optical gap
$\tilde{\Delta}=\Delta\gamma_1/\sqrt{\gamma_1^2+\Delta^2}$
that lies at a finite momentum ${\bf k}$ at the bottom of the mexican hat
 (Fig. \ref{f-chijS}b).
The reduced dimensionality of the electronic dispersion
in these points gives rise to a singular behavior in the density
of states, which is reflected in a corresponding behavior
in the particle-hole excitations in $\xi_{\rm S}''(\omega)$.
Such peaked structure in $\xi_{\rm S}''(\omega)$ has important
consequences also on the real part of $\xi_{\rm S}(\omega)$,
resulting in a strong peak at $\omega \approx \tilde{\Delta}$.
When $\tilde{\Delta} \approx  0.2$ eV ($\Delta \approx  0.233$ eV)
both $\xi_{\rm S}'(\omega)$ and $\xi_{\rm S}''(\omega)$
are peaked at $\omega \approx \omega_{\rm S}$
and the phonon strength is expected to be
strongly enhanced.

\begin{figure}[t]
\begin{center}
\includegraphics[scale=0.35,clip=]{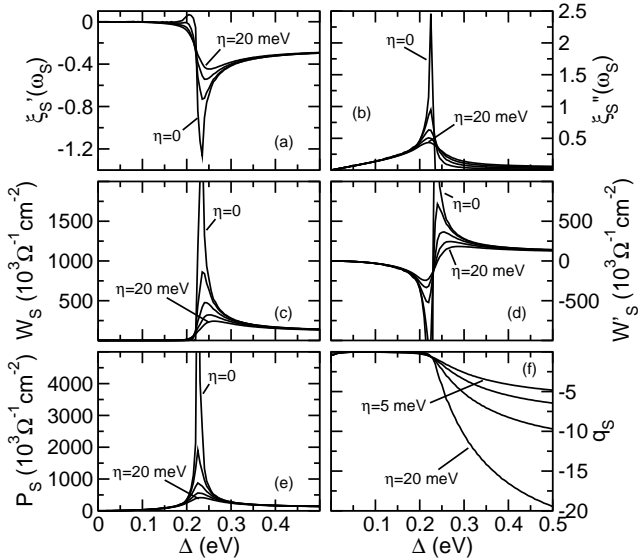}
\end{center}
\caption{Dimensionless quantities
$\xi'(\omega_{\rm S})$, $\xi''(\omega_{\rm S})$,
as well as the spectral weights $W_{\rm S}$, $W_{\rm S}'$,
$P_{\rm S}$ and the Fano factor $q_{\rm S}$
as functions of the gap $\Delta$ at $\mu=0$
and for different values of the damping
$\eta= 0, 5, 10, 15, 20$ meV.}
\label{f-gap_mu0}
\end{figure}

This trend, as well as the dependence on $\Delta$
of all the optical properties of the S phonon peak,
are shown in Figs. \ref{f-gap_mu0}.
For $\tilde{\Delta} \lesssim \omega_{\rm S} \approx 0.2$ eV
the phonon strength $P_{\rm S}$ is increasing
with $\Delta$ signalizing the switch-on of the
symmetric $E_g$ mode. Such increase of the strength is mainly
driven by $\xi''(\omega_{\rm S})$ while $\xi'(\omega_{\rm S})\approx 0$.
This results in a {\em negative} peak
where $W_{\rm S} \approx 0$ and a negative integrate area
$W_{\rm S}' < 0$. The vanishing of $\xi'(\omega_{\rm S})$ results
also in a  Fano factor $q_{\rm S}\approx 0$, which indeed
corresponds to a negative symmetric shape.

\subsection{Doped case $n\neq 0$, $\Delta \neq 0$}

Unless using a double-gate device,\cite{tang} where the top and bottom
gates are tuned to set the doping or the gap to zero,
in single gated devices
the gating induces at the same time
an inversion symmetry breaking and a finite doping.
In this most common case, both the $E_u$ and $E_g$ modes
are simultaneously IR active.
In order to investigate theoretically the optical properties
of the phonon resonances, one has to employ
the full Eq. (\ref{chirice}), where both modes are presents.
The phonon spectral properties are then
much more complex than in the $\Delta=0$ case.
Not only we have different phonon channels contributing
to the total features, but also the phonon propagator
of each channel [e.g. the $D_{\rm SS}(\omega)$ propagator],
under particular conditions,
can develop a double-pole structure, as discussed in Refs. \onlinecite{malard,Ando_09,gava,yan3}
in the context of the Raman spectroscopy.

In general, we can attribute a different role
to the several quantities appearing in Eq. (\ref{chirice}).
Here the mixed response functions $\chi_{j\nu}$ are mainly
responsible for the magnitude and the Fano lineshape of
the phonon features, while the pole structure of the phonon
propagators $D_{\nu\nu'}$, is associated
with the frequencies of the phonon resonances and to their
linewidth. Keeping in mind this distinction,
and since the frequency structure of $\chi_{j\nu}(\omega)$
varies over electronic energy scales while
the phonon self-energy gives rise
to a splitting of the S and A mode frequencies of few cm$^{-1}$,
we can in a very good approximation evaluate
the functions $\chi_{j\nu}(\omega)$ at the bare phonon frequency
$\omega \approx \omega_0$.
In this context, the relevance of each phonon mode
in the infrared spectroscopy is ruled by the characteristic
phonon strength of the corresponding channel, i.e.
$P_{\rm A}=\pi |\chi_{j{\rm A}}(\omega_0)|^2/\omega_0 V$
for the antisymmetric $E_u$ mode, associated
with the first term in the r.h.s. of Eq. (\ref{chirice}),
and
$P_{\rm S}=\pi |\chi_{j{\rm S}}(\omega_0)|^2/\omega_0 V$
for the symmetric $E_g$ mode, associated
with the second term in the r.h.s. of Eq. (\ref{chirice}).
It is also possible to define a mixed channel,
related with the third term in the r.h.s. of Eq. (\ref{chirice}),
characterized by a phonon strength
$P_{\rm AS}=\pi  \sqrt{|\chi_{j{\rm A}}(\omega_0)
\chi_{{\rm S}^\dagger j}(\omega_0)|}/\omega_0 V$.
This channel is however quite weak
in most cases,\cite{cbk}
and we neglect it in the following discussion.

\begin{figure}[t]
\begin{center}
\includegraphics[angle=-90,scale=0.38,clip=]{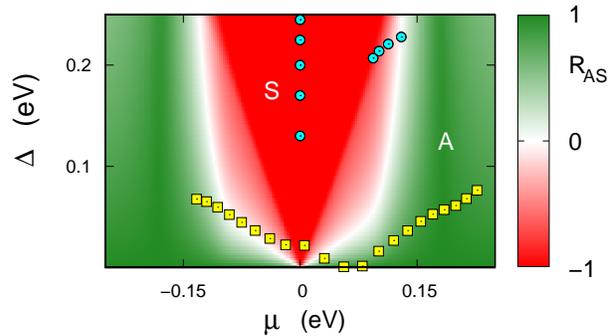}
\end{center}
\caption{Phase diagram of the
relative relevance of different phonon channels
$R_{\rm AS}=(P_{\rm A}-P_{\rm S})/(P_{\rm A}+P_{\rm S})$.
In the green region optical features associated
with the antisymmetric $E_u$ mode are dominant,
in the red one the dominant mode
is the symmetric $E_g$ one.
Also shown here is the location in the $\mu$-$\Delta$ space
of the experimental available data [Refs. \onlinecite{noi} (squares),
\onlinecite{tang} (circles)].
Adapted from Ref. \onlinecite{cbk}.
}
\label{f-map}
\end{figure}

The comparison of the phonon strengths associated
with different A and S phonon modes
is shown in Fig. \ref{f-map},
where we plot
the relative intensity
$R_{\rm AS}=(P_{\rm A}-P_{\rm S})/(P_{\rm A}+P_{\rm S})$
as a function of
the chemical potential $\mu$ (doping)
and the band gap $\Delta$ induced by the vertical electric field.
As we discussed above,
$P_{\rm A}$ is essentially driven by the doping,
whereas $P_{\rm S}$ is induced by the gap $\Delta$.
According to the relative position in the $\mu$ vs. $\Delta$
phase diagram, we can thus predict one mode
to be dominant with respect to the other one.
A ``phonon switching'',
namely the switch from the dominance of one
phonon mode to another mode,
is possible as a function of the gate voltage.
In Fig. \ref{f-map} we also show
the location in the $\mu$-$\Delta$ space
of the experimental available data from Refs. \onlinecite{noi,tang}.
While the data of Ref. \onlinecite{tang}, in the double-gated device,
were mainly collected along the neutral line $\mu=0$,
the optical conductivity measurements of Ref. \onlinecite{noi}
in the single-gate geometry span a much wider
region, going from regions where the antisymmetric
mode $E_u$ is expected to dominate to regions where the optical features can
be attributed mainly to the symmetric mode $E_g$.
The actual evidence of such theoretically-predicted phonon
switching in the experimental data was discussed
in Ref. \onlinecite{cbk}, where we refer the reader
for more details. 

One should underline that
the phonon switching is not directly related
to the appearing of a double-peak structure
in the optical conductivity, as observed
in Raman spectroscopy.\cite{malard,Ando_09,gava,yan3}
What we are describing here
is the dominant intensity of one phonon with respect
to the other one, in other words, which phonon mode
is most coupled to the light.
This information is encoded in the mixed response function
$\chi_{j\nu}$ which describes the coupling of the
light with the mode $\nu$.
In this respect, as shown in Fig. \ref{f-map},
in bilayer graphene the two channels are 
to a good extent mutually exclusive.
A different matter is the possibility, within
a given phonon channel, to develop a two-peak phenomenology.
This issue is related to the presence
of a large off-diagonal phonon self-energy $\chi_{\rm A,S}$
of the same order of the diagonal
phonon self-energies $\chi_{\rm A,A}$, $\chi_{\rm S,S}$
in the $2 \times 2$ space of the phonon modes $\nu=$A, S.
On this respect,
since the phonon self-energy is a quantity which
is shared in both Raman and IR spectroscopy,
we expect that phonon features in optical conductivity
can develop double-peak structures
in the same regions where the Raman spectroscopy sees them.
Things are however more complicated
in the optical conductivity case since, as we show
in Section \ref{s-raman}, Raman spectroscopy in bilayer graphene
is dominated only by the S channel.
On the other hand, in the optical conductivity,
we expect that the condition $\chi_{\rm A,S}\neq 0$,
ruling the double-peak features, would arise
in the same phase space where the
mixed channel $P_{\rm AS}$ [second line in Eq. (\ref{chirice})]
is of the order of $P_{\rm A}$, $P_{\rm S}$.
In this situation spectral interferences between
the different channels can occur making the scenario
more complex than in Raman spectroscopy.

\section{Full tight-binding model}
\label{s-TB}

The simplified model considering only the leading tight binding
terms $\gamma_0$ and $\gamma_1$ has some significant limitations.
For example, within this model
the intensity of the IR active $E_u$
drops exactly to zero
in the undoped limit $\mu\rightarrow 0$.
The generalization of this model to
bulk graphite would predict thus no IR activity at all,
despite the clear evidence of a phonon resonance
reported already in the 70s.\cite{nemanich,dresselhaus}
As a matter of fact, the evidence of such phonon intensity
in graphite was reconciled with the charged-phonon theory
in Ref. \onlinecite{manzardo} where it was shown that
the inclusion of higher order tight-binding terms,
in particular of the ones breaking the particle-hole symmetry,
is responsible for 
the observed phonon activity.
It is thus interesting to investigate to which extent
these higher-order tight-binding terms
can affect
the results for the bilayer graphene.

We address this issue by including explicitly
in the Hamiltonian of bilayer graphene
the higher-order tight-binding hoppings
$\gamma_3$, $\gamma_4$ as well as the crystal field
$\delta$ which differentiate the atoms B1, A2
from the atoms A1, B2.
Close to the K point,
we can write:
\begin{eqnarray}
\label{hamiltonianTB}
\hat{H}_{\bf k}=
\left(
\begin{array}{cccc}
\Delta/2 & v {\bf \pi}_- & v_4 {\bf \pi}_-&
v_3 {\bf \pi}_+\\
 v {\bf \pi}_+ & \delta+\Delta/2 &
\gamma_1  & v_4 {\bf \pi}_-\\
v_4 {\bf \pi}_+ & \gamma_1 &\delta -\Delta/2 &
v {\bf \pi}_- \\
 v_3 {\bf \pi}_-& v_4 {\bf \pi}_+& v {\bf \pi}_+ & -\Delta/2
\end{array}
\right),
\end{eqnarray}
where $v_i=\gamma_i/\gamma_0$.
We consider here
typical values of $\gamma_3=0.29$ eV,
$\gamma_4=-0.13$ eV and $\delta=0.022$ eV.\cite{pp}
We evaluate consequently also the current operator
\begin{eqnarray}
\label{currTB}
\hat{j}_{{\bf k},y}
&=&-\frac{e}{\hbar}
\frac{d}{dk_y}\hat{H_{\bf k}}
\nonumber\\
&=&
-e v \hat{I}(\hat{\sigma}_y)
-e v_4 \hat{\sigma}_y(\hat{I})
+e v_3 \frac{
\hat{\sigma}_x(\hat{\sigma}_y)+\hat{\sigma}_y(\hat{\sigma}_x)
}{2},
\end{eqnarray}
and the electron-phonon scattering matrices
\begin{eqnarray}
\label{vaTB}
\hat{V}_{\rm A}
&=&
i g \hat \sigma_z(\hat \sigma_x)
+
i g_4 \hat{\sigma}_x(\hat{\sigma}_z),
\\
\hat{V}_{\rm S}
&=&
i g \hat I(\hat \sigma_x)
-
i g_3 \frac{
\hat{\sigma}_x(\hat{\sigma}_x)-\hat{\sigma}_y(\hat{\sigma}_y)
}{2}.
\end{eqnarray}
The terms $g_3$, $g_4$ represent the electron-phonon coupling
associated respectively
with the hopping terms $\gamma_3$ and $\gamma_4$ and
they can be related to the corresponding deformation potentials
which have been recently evaluated by DFT calculations.\cite{cp}
We get namely $g_3=0.033$ eV and $g_4=0.018$ eV.

\begin{figure}[t]
\begin{center}
\includegraphics[scale=0.35,clip=]{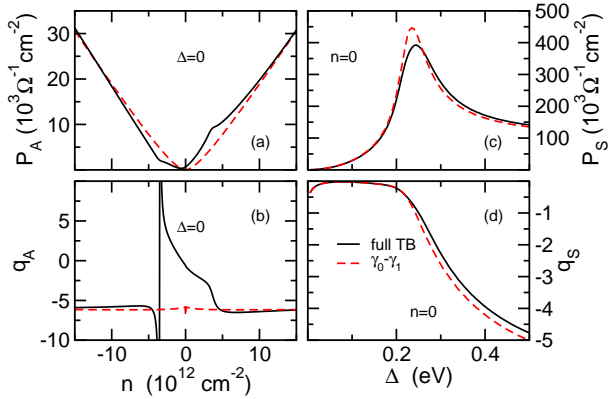}
\end{center}
\caption{Phonon strength $P_{\rm A}$ (panel a)
and Fano parameter $q_{\rm A}$  (panel b)
for the $E_u$ mode
as functions of doping $n$ for $\Delta=0$
including all the relevant tight-binding parameters
(black solid line) and using the simple $\gamma_0$-$\gamma_1$
model of Section 4 (red dashed line), with $T=10$ K and $\eta=20$ meV.
Panels c and d: same quantities $P_{\rm S}$, $q_{\rm S}$
for the $E_g$ mode at $n=0$ as functions
of the gap $\Delta$.}
\label{f-tb}
\end{figure}

As two representative limits, we
show in Fig. \ref{f-tb}a,b the phonon strength $P_{\rm A}$
and the Fano parameter $q_{\rm A}$ for the $E_u$ mode
in the ungapped case as functions of the doping $n$,
and in Fig. \ref{f-tb}c,d the phonon strength $P_{\rm S}$
and the Fano parameter $q_{\rm S}$ for the $E_g$ mode
in the undoped case as functions of the
electric field parametrized by the gap $\Delta$
at the K point.
As found for bulk graphite in Ref. \onlinecite{manzardo},
a finite phonon strength $P_{\rm A}\approx 0.62 \times 10^3$ $\Omega^{-1}$
cm$^{-2}$
(weakly visible on the scale of Fig. \ref{f-tb}a)
is now triggered by the higher-order tight-binding terms
at the neutral point.
However,
apart from a slight asymmetry for electron and hole doping,
the inclusion of such tight-binding terms does not
change qualitatively the results at finite doping,
with a roughly linear
increasing of the phonon strength $P_{\rm A}$
as function of $|n|$.
In addition, such residual phonon strength $P_{\rm A}$
at the neutrality point is easily 
overwhelmed in gated systems by the presence
of the symmetric $E_g$ mode with $P_{\rm S} \gg P_{\rm A}$
(Fig. \ref{f-tb}c).
The $\Delta$ dependence of the optical
properties of such mode (phonon strength $P_{\rm S}$,
Fano parameter $q_{\rm S}$) is also barely affected
by the presence of the higher-order tight-binding terms,
so that we can conclude that the scenario presented
in Sections \ref{s-nogap}, \ref{s-gap} is generally
robust in realistic materials
against the inclusion of higher order tight-binding terms,
except in the case of
ungated and undoped bilayer graphene,
as it could be relevant in suspended samples.

\section{Charged phonon theory for the Raman response}
\label{s-raman}

In the previous Sections we have outlined in details how
a quantitative implementation of the charged-phonon theory
gives rise in the infrared optical spectroscopy of bilayer graphene
to a phonon intensity and to a asymmetric Fano-like shape
for the A mode as well as for
the S mode once this latter is turned on
by the interlayer potential.
In particular, such analysis shows that
for gate potential $V_g$ close to the charged neutrality point,
the negative phonon peak can been attributed to the S mode
with a phonon activity strongly dependent on the gate voltage $V_g$
and with a strong Fano character with $q_{\rm S}\approx 0$,
induced by the interference with the low-energy
2-3 interband particle-hole excitations.

One can
wonder why the same S mode, under the same conditions
(namely tuning charge concentration and bandgap $\Delta$) does not present
in the Raman response
any significant Fano asymmetry and strong dependence
of the phonon intensity upon $V_g$.\cite{yan2,basko1,casiraghi}
In this Section we show that this different behavior
for the infrared and Raman spectroscopy can be naturally
explained within the context of the charged-phonon theory.

In our analysis we focus on the possible changes
in the Raman features of the phonon resonances at $\sim 1590$ cm$^{-1}$ as
functions the charge doping and the bandgap,
i.e. quantities that typically affect only the low-energy excitations
close to the Fermi level.
We do not address the possible dependence
of the phonon intensity on the external laser energy.\cite{basko2,basko3}
We also assume a relatively small gate-induced doping,
so that the chemical potential is much less than the laser energy,
in a region where the absolute Raman intensity of the phonon peak
at $\sim 1590$ cm$^{-1}$ is constant.\cite{chen}
Within this context,
we employ the
effective mass approximation,\cite{devereaux} when
the only relevant quantity is the Raman shift
$\omega=\omega_{\rm in}-\omega_{\rm out}$,
of the order of the phonon energy, $\omega\approx\omega_0$.
Refs. \onlinecite{basko2,basko3} showed also how the main electronic
transitions responsible for the phonon intensity were associated
with high energy processes, of the order of the  laser energy $\omega_{\rm in}\approx 1$ eV
or of the order of the $\pi$-bandwidth $W \approx 6-7$ eV, in any case much
larger than the phonon energy scale.
As we shall see,
we will recover this result in our simplified scheme and we will show
how it can explain the different phenomenology of Raman
spectroscopy with respect to the infrared spectroscopy.
For simplicity we
consider unpolarized isotropic Raman scattering.
The Raman intensity $I(\omega)$ can be related to the imaginary part
of the Raman response function,\cite{devereaux} namely
\begin{eqnarray}
I(\omega)
=
-\frac{1}{\pi}
[1+b(\omega/T)]\mbox{Im}\chi_{\rm RM}(\omega),
\end{eqnarray}
where $b(x)=1/[\exp(x)-1]$ is the Bose-Einstein function and
$\chi_{\rm RM}(\omega)$ is the analytical continuation
to the real frequency axis of the Raman response function
$\chi_{\rm RM}(i\omega_m)$, whose
explicit definition and derivation are given
in Appendix \ref{app-raman}.
Just as in the case of the optical conductivity,
the irreducible part of $\chi_{\rm RM}(\omega)$ provides
the electronic Raman background, which is
proportional
to the current-current response function involved
in the optical conductivity.
In a similar way, within the charged-phonon theory,
the bubble diagrams mediated by a phonon propagator,
as in Fig. \ref{f-diagrams}b,
are associated with the onset of the phonon peaks
in the Raman response.
In particular, we can write, for $\Delta \neq 0$,
\begin{eqnarray}
\Delta\chi_{\rm RM}(\omega)
&=&
\chi_{\gamma{\rm S}}(\omega)D_{\rm SS}(\omega)
\chi_{{\rm S}^\dagger \gamma}(\omega)
\nonumber\\
&&
+\chi_{\gamma{\rm A}}(\omega)D_{\rm AA}(\omega)
\chi_{{\rm A}^\dagger \gamma}(\omega)
\nonumber\\
&&
+
\left[
\chi_{\gamma{\rm S}}(\omega)D_{\rm SA}(\omega)
\chi_{{\rm A}^\dagger \gamma}(\omega)
+\mbox{h.c.}
\right],
\label{riceRm}
\end{eqnarray}
where $\chi_{\gamma\nu}(\omega)$ are Raman mixed response
functions involving one Raman vertex $\gamma$,
described by the effective mass approximation, and one
electron-phonon scattering operator.
Explicit expressions for the Raman vertices for different
polarizations are given
in Appendix \ref{app-raman}.
It is worthwhile to point out that
close to the K point
the Raman vertex scattering operator for the $xx$ polarization
involved in Eq. (\ref{riceRm}) reads
\begin{eqnarray}
\hat{\gamma}_{\bf k}^{xx}\propto \hat{I}(\hat{\sigma}_x)
\propto i \hat{V}_{\rm S}.
\label{rmph}
\end{eqnarray}
This observation permits us to relate the Raman spectroscopy
in the effective-mass approximation to the response
functions previously discussed.
In particular,
using the relation (\ref{rmph}), we can relate
the Raman mixed response functions to the phonon self-energy.
We obtain for instance
$\chi_{\gamma{\rm S}}(\omega)\propto \chi_{\rm SS}(\omega)$
and $\chi_{\gamma{\rm A}}(\omega)\propto \chi_{\rm SA}(\omega)$.
Just as in the infrared response, also in this case
the Raman spectral strength of each mode will be
$P_\nu^{\rm Raman}\propto |\chi_{\gamma\nu}(\omega_\nu)|^2$
and the Fano asymmetry factor
$q_\nu^{\rm Raman}=-\chi_{\gamma\nu}'(\omega_\nu)
/\chi''_{\gamma\nu}(\omega_\nu)$.

A crucial difference with respect to the infrared spectroscopy
is the different dependence of the Raman mixed
response functions $\chi_{\gamma\nu}$ on the high-energy interband transitions
when compared to the infrared response functions $\chi_{j\nu}$.
One should keep in mind that the linear Dirac-like dispersion
$\epsilon=\hbar v|{\bf k}|$ of the single-layer graphene
extends only up to 3-5 eVs.
However,
as we saw in Section \ref{s-nogap},
in the clean limit and in the absence of the gap $\Delta$ 
the imaginary part of $\chi_{j{\rm A}}(\omega)$,
and hence also the real part due to the Kramers-Kronig relations,
are uniquely determined by low-energy interband transitions
$\gamma \le |\omega| \le \gamma +2|\mu|$,
so that the high-energy cut-off $E_c$ does not play any role.
One can see that such results hold true also for
$\chi_{j{\rm S}}(\omega)$ and for
generic finite $\eta$ and $\Delta$, in a sense that
that the imaginary part of the infrared mixed response functions
$\chi_{j\nu}(\omega)$ is not divergent
for $\omega \rightarrow \infty$
so that the high-energy cut-off $E_c$ can be safely set to infinity.

The situation is different when the Raman response function is considered.
In this case, using Eq. (\ref{rmph}), we get
$\chi_{\gamma{\rm S}}\propto \chi_{\rm SS}$
and $\chi_{\gamma{\rm A}}\propto \chi_{\rm SA}$.
The function $\chi_{\rm SS}(\omega)$, which corresponds
to the phonon self-energy for the symmetric $E_g$ mode,
has been widely analyzed in the literature.\cite{ando2L,Ando_09,gava}
In particular, its imaginary part increases
linearly at high energy, $\chi_{\rm SS}''(\omega)\sim \omega$.
As a consequence,
from the Kramers-Kronig relations,
the magnitude of the real part of $\chi_{\rm SS}(\omega)$
is dominated by the high-energy processes, {\em i.e.}
by high-energy cut-off $E_c$, $\chi_{\rm SS}'(\omega) \sim E_c$.
On a physical ground, this cut-off $E_c$
can be identified with the $\pi$-bandwidth $W$
or with the highest energy $\omega_{\rm in}$ above which
the effective mass approximation breaks down.

From a careful inspection, one can see
that both the real and imaginary parts of 
almost all the mixed response functions 
$\chi_{\gamma,\nu}$
are only weakly dependent on the high energy cut-off $E_c$.
The only exception is the real part
of the Raman coupling with the S mode,
Re$\chi_{\gamma,{\rm S}}$, which
scales with $E_c$.
Since this is
the highest energy scale involved in the system, we have
$\chi_{\gamma{\rm S}}' \gg
\chi_{\gamma{\rm S}}''$, $\chi_{\gamma{\rm A}}'$,
$\chi_{\gamma{\rm A}}''$.
As a consequence we see that the
phonon Raman spectroscopy is dominated by
direct coupling of the Raman scattering operator with the $E_g$
symmetric mode [first line of Eq. (\ref{riceRm})],
whereas other channels involving the $E_u$ are marginal,
independently of the charge doping level or of the
interlayer potential difference $\Delta$.
Other interesting consequences:
$i$) since $\chi_{\gamma{\rm S}}' \gg
\chi_{\gamma{\rm S}}''$, 
the Fano factor for the dominant S channel results
$|q_{\rm S}^{\rm Raman}|=|\chi_{\gamma{\rm S}}'(\omega_{\rm S})
/\chi_{\gamma{\rm S}}''(\omega_{\rm S})| \gg 1$ and
the Raman phonon peaks are expected to be positive
and symmetric, in agreement with the experiments;
$ii$) the magnitude of the Raman phonon strength
$P_{\rm S}^{\rm Raman} \propto [\chi_{\gamma{\rm S}}']^2$
is mainly dominated by the energy cut-off $E_c$, so that
it does not depend significantly on the low-energy features
related to the charge doping or to the opening of the gap $\Delta$,
also in agreement with the experimental observations.

Note that the fact that the phonon Raman response is dominated
by the direct coupling of polarizability to the S mode does not exclude
that a double-peak structure could be observed.
Indeed, as we saw in Section \ref{s-gap},
under suitable conditions the
phonon propagator $D_{\rm SS}(\omega)$ itself, as well
as $D_{\rm AA}(\omega)$, can develop a double-peak structure
as a consequence of the phonon-modes hybridization triggered
by the mixed phonon self-energy $\chi_{\rm AS}$.\cite{malard,Ando_09,gava,yan3}

\section{Conclusions}
\label{s-conclusions}

In this paper we have provided a comprehensive derivation
of the charged-phonon theory applied to investigate the optical properties
of the phonon peaks in the optical conductivity of bilayer graphene.
The origin of the phonon activity and its relation with the occurrence
of a Fano effect have been elucidated. The dependence of these properties
on the tunable microscopical parameters, i.e. the doping and the
bandgap induced by an external gate voltage, has been discussed in detail.
We have also compared the charged-phonon theory
in the infrared and Raman spectroscopy, accounting for the
different phenomenology of the phonon peaks in these two
different optical probes.

The theory presented here provides a suitable tool to characterize
quantitatively bilayer graphene
in terms of the intensity and Fano asymmetry of the infrared
phonon peaks.
Further future developments of the present analysis could
investigate the dependence of the phonon optical properties on the different
symmetry breakings,\cite{edv,vafek,zh,levitov,mucha11,rutter,pono,abergel}
in order to provide a fingerprint
for the possible underlying instabilities.
The microscopical characterization of the phonon optical properties
sheds also light
on the underlying physics of the electron-phonon interaction.
While the present work was focused on bilayer graphene,
as the simplest graphitic system with infrared active modes,
the theory presented  here can be generalized in a straightforward way
to multilayer graphene,\cite{li} and to bulk
graphite.\cite{manzardo}

The general scheme discussed in
this work to investigate within a microscopic many-body approach the
Fano effect can be applied, with the due modifications, also to different
classes of materials. One remarkable example is provided by  
layered systems with different atomic species
in the units cell, for instance MoS$_2$, Bi$_2$Se$_3$,
where phonon anomalies have been detected in
infrared and Raman spectroscopy.\cite{lee,laforge} A second
interesting case is the As phonon mode in the pnictide 
Ba(Fe$_{1-x}$Co$_x$)$_2$As$_2$, which displays a concomitant
presence of intensity variation and asymmetry as a function of doping
and/or temperature, in particular across the magnetostructural
transition.\cite{gallais} In all these cases the understanding of the phonon
anomalies can shed new light on the underlying bulk electronic
structure. In this sense phonon spectroscopy 
can represent a powerful and alternative tool to investigate electronic
excitations also in correlated materials, provided that a correct implementation of the Fano-Rice
theory is used to relate the phonon and electronic response, taking
into account the microscopic selection rules. 

\section*{Acknowledgements}

E.C. acknowledges the Marie Curie grant PIEF-GA-2009-251904.
The work of A.B.K. was supported by the grant No.200020-130093
of the Swiss National Science Foundation (SNSF).

\appendix
\section{Analytical results for the charged-phonon theory
of IR response in ungapped bilayer graphene}
\label{app-chijA}

In this Appendix we provide the analyical derivation
of the optical properties of the A mode in the absence
of electrostatic bias between the layers,
$\Delta=0$.
To this aim it is convenient, for the simple tight-binding
model with $\gamma_0$-$\gamma_1$ considered, 
and in the linear expansion close to the K point, to employ
the cylindric coordinates, where the Hamiltonian
reads:
\begin{eqnarray}
\label{hamiltonian_aa}
\hat{H}_{\bf k}=
\left(
\begin{array}{cccc}
\Delta/2 & \epsilon_k^- & 0 & 0\\
 \epsilon_k^+ & \Delta/2 & \gamma_1  & 0\\
0 & \gamma_1 & -\Delta/2 & \epsilon_k^- \\
0 & 0 & \epsilon_k^+ & -\Delta/2
\end{array}
\right),
\end{eqnarray}
where $\epsilon_k^\pm=\epsilon_k \mbox{e}^{\pm i\theta_k}$
$\epsilon=\hbar v |k|$, and $\theta_k=\arctan(k_y/k_x)$.

The Hamiltonian is diagonalized by the transformation:
\begin{eqnarray}
\tilde{\hat{H}}_k
&=&
\hat{M}_{\bf k}^{-1} \hat{H}_{\bf k}\hat{M}_{\bf k}
=
\hat{E}_k,
\end{eqnarray}
where
\begin{eqnarray}
\hat{E}_k
&=&
\left(
\begin{array}{cccc}
E_{1k} & 0 & 0 & 0 \\
0 & E_{2k} & 0 & 0 \\
0 & 0 & E_{3k} & 0 \\
0 & 0 & 0 & E_{4k} \\
\end{array}
\right),
\label{Ematr}
\end{eqnarray}
(band labels according Fig. \ref{f-basis}b),
\begin{eqnarray}
\hat{M}_{\bf k}
&=&
\hat{R}_{\theta_k}^{-1}\hat{F}_k,
\end{eqnarray}
\begin{eqnarray}
\hat{R}_\theta
&=&
\left(
\begin{array}{cccc}
\mbox{e}^{i\theta_k} & 0 & 0 & 0 \\
0 & 1 & 0 & 0 \\
0 & 0 & 1 & 0 \\
0 & 0 & 0 & \mbox{e}^{-i\theta_k} \\
\end{array}
\right),
\label{rmatr}
\end{eqnarray}
\begin{eqnarray}
\hat{F}_k
&=&
\frac{1}{\sqrt{2}}
\left(
\begin{array}{cccc}
-s_k & -c_k & c_k & s_k \\
c_k & s_k & s_k & c_k \\
-c_k & s_k & -s_k & c_k \\
s_k & -c_k & -c_k & s_k
\end{array}
\right).
\label{Fmatr}
\end{eqnarray}
Here
\begin{eqnarray}
s_k
&=&
\frac{1}{\sqrt{2}}
\sqrt{
1-\frac{\gamma_1}{2\sqrt{\gamma_1^2/4+\epsilon_k^2}}
},
\\
c_k
&=&
\frac{1}{\sqrt{2}}
\sqrt{
1+\frac{\gamma_1}{2\sqrt{\gamma_1^2/4+\epsilon_k^2}}
},
\end{eqnarray}
so that
\begin{eqnarray}
2s_kc_k
&=&
\frac{|\epsilon_k|}{\sqrt{\gamma_1^2/4+\epsilon_k^2}}.
\label{cksk}
\end{eqnarray}

In the diagonalized basis we have:
\begin{eqnarray}
\tilde{\hat{j}}_{{\bf k},y}
&=&
\hat{M}_{\bf k}^{-1} \hat{j}_{{\bf k},y}\hat{M}_{\bf k}
\nonumber\\
&=&
-ev
\left(
\begin{array}{cccc}
-S_k s_\theta & -iC_k c_\theta &
C_k s_\theta & iS_k c_\theta \\
iC_k c_\theta & -S_k s_\theta &
iS_k c_\theta & -C_k  s_\theta \\
C_k s_\theta & -iS_k c_\theta &
S_k s_\theta & -iC_k c_\theta \\
-iS_k c_\theta & -C_k s_\theta &
iC_k c_\theta & S_k s_\theta
\end{array}
\right),
\label{ymatr}
\end{eqnarray}
where $S_k=2s_kc_k$, $C_k=c_k^2-s_k^2$, $s_\theta=\sin\theta$,
$c_\theta=\cos\theta$.

In a similar way we obtain:
\begin{eqnarray}
\tilde{\hat{V}}_{{\bf k},{\rm A}}
&=&
i g 
\left(
\begin{array}{cccc}
0 & -c_\theta & -is_\theta & 0 \\
-c_\theta & 0 & 0 & -is_\theta \\
is_\theta & 0 & 0 & c_\theta \\
0 & is_\theta & c_\theta & 0
\end{array}
\right),
\label{lambdamatr}
\\
\tilde{\hat{V}}_{{\bf k},{\rm S}}
&=&
i g 
\left(
\begin{array}{cccc}
-S_k c_\theta & iC_k s_\theta & C_k c_\theta & -iS_k s_\theta \\
- iC_k s_\theta & -S_k c_\theta & -iS_k s_\theta & -C_k c_\theta \\
C_k c_\theta & iS_k s_\theta & S_k\ c_\theta & iC_k s_\theta \\
iS_k s_\theta & -C_k c_\theta& -iC_k s_\theta & S_k c_\theta
\end{array}
\right).
\label{lambdamatrRaman}
\end{eqnarray}

Note that the interband transitions between bands $1$-$4$ and $2$-$3$
are missing in the electron-phonon matrix (\ref{lambdamatr})
for the A mode, so that these interband transitions
will not be operative in
any response function involving
such electron-phonon coupling, like for instance
the mixed response function for the infrared activity
as well as the phonon self-energy.

The mixed response function reads in the diagonalized basis as
\begin{eqnarray}
\chi_{j\nu}(i\omega_m)
&=&
N_s N_v \frac{T}{N}\sum_{{\bf k},n}
\mbox{\rm Tr}\Big[
\tilde{\hat{j}}_{{\bf k},y}
\hat{g}(k,i\omega_n+i\omega_m)
\nonumber\\
&&
\times
\tilde{\hat{V}}_{{\bf k},\nu}
\hat{g}(k,i\omega_n)
\Big],
\end{eqnarray}
where $\hat{g}(k,i\omega_n)=
1/[(i\hbar\omega_n+\mu)\hat{I}-\tilde{\hat{H}}_k]$,
is the electronic Green's function in the diagonalized basis.

Using now Eqs. (\ref{Ematr}), (\ref{ymatr}),
(\ref{lambdamatr}), (\ref{lambdamatrRaman}),
we can easily obtain:
\begin{eqnarray}
\chi_{j\nu}(i\omega_m)
&=&
\frac{1}{N}\sum_{{\bf k},\alpha,\beta}
C^{\alpha\beta}_{j\nu,{\bf k}}
\Pi^{\alpha\beta}_k(i\omega_m),
\end{eqnarray}
where
\begin{eqnarray}
C^{\alpha\beta}_{j\nu,{\bf k}}
&=&
N_s N_v
(\tilde{\hat{j}}_{{\bf k},y})_{\alpha\beta}
(\tilde{\hat{V}}_{{\bf k},\nu})_{\beta\alpha},
\end{eqnarray}
and 
\begin{eqnarray}
\Pi^{\alpha\beta}_k(i\omega_m)
&=&
\frac{f(E_{k,\alpha}-\mu)-f(E_{k,\beta}-\mu)}
{E_{k,\alpha}-E_{k,\beta}+\hbar i\omega_m}.
\end{eqnarray}
Note that the function $\pi^{\alpha\beta}_k(i\omega_m)$
does not depend, in cylindrical coordinates, on the angle
$\theta_k$ but only on the momentum modulus $k=|{\bf k}|$.
Writing $\sum_{\bf k}=\int 2\pi kdk \int d\theta_k/2\pi$,
the quantity $C^{\alpha\beta}_{j\nu,{\bf k}}$
can be replaced thus with its average over the angle $\theta_k$,
$C^{\alpha\beta}_{j\nu,{\bf k}} \rightarrow
C^{\alpha\beta}_{j\nu,k}
=
\langle C^{\alpha\beta}_{j\nu,{\bf k}}\rangle_{\theta_k}$.
Using (\ref{lambdamatr}), (\ref{lambdamatrRaman}),
it is now easy to see that
$C^{\alpha\beta}_{j{\rm S},k}$
averages out, whereas
$|C^{\alpha\beta}_{j{\rm A},k}|=ge vN_s N_v
\gamma_1/4\sqrt{(\hbar v k)^2+\gamma_1^2}$, which recovers
Eq. (\ref{cfunc}). The relative change of sign between
different interband contributions stems from
$(\tilde{\hat{j}}_{{\bf k},y})_{\alpha\beta}
(\tilde{\hat{V}}_{{\bf k},\nu})_{\beta\alpha}
=-(\tilde{\hat{j}}_{{\bf k},y})_{\beta\alpha}
(\tilde{\hat{V}}_{{\bf k},\nu})_{\alpha\beta}$

\section{Charged-phonon theory for Raman spectroscopy}
\label{app-raman}

In this Appendix we provide a brief derivation of the
charged-phonon theory as applied to the case
of Raman spectroscopy within the effective mass approximation.
To this end it is useful to recall the tight-binding Hamiltonian
which can be written as
\begin{eqnarray}
\hat{H}_{\bf p}
&=&
\left(
\begin{array}{cccc}
\Delta/2 & \gamma_0f_{\bf  p}^* & 0 & 0 \\
\gamma_0f_{\bf p} & \Delta/2 & \gamma_1 & 0 \\
0 & \gamma_1 & -\Delta/2  & \gamma_0 f_{\bf p}^* \\
0 & 0 & \gamma_0 f_{\bf p} & -\Delta/2 \\
\end{array}
\right),
\label{aham}
\end{eqnarray}
where
\begin{eqnarray}
f_{\bf p}
&=&
\mbox{e}^{-ip_xa/\sqrt{3}}
+2\mbox{e}^{ip_xa/2\sqrt{3}}\cos(p_ya/2).
\end{eqnarray}

Close to the K=$(4\pi/3a,0)$ point
we can write ${\bf p}=\mbox{K}+{\bf k}$, and by expanding for small
${\bf k}$ we obtain (\ref{hamiltonian}).
The Raman vertex $\hat{\gamma}(\phi,\phi')$
for a particular polarization geometry
can be now defined as
$\gamma(\phi,\phi')=(1/N)\sum_{{\bf k},\sigma}
\Psi_{{\bf k},\sigma}^\dagger \hat{\gamma}_{\bf k}(\phi,\phi')
\Psi_{{\bf k},\sigma}$
where
\begin{eqnarray}
\hat{\gamma}_{\bf k}
&=&
({\bf e}_{\rm i}\cdot \nabla_{\bf k})
({\bf e}_{\rm o}\cdot \nabla_{\bf k})
\hat{H}_{\bf k}
\nonumber\\
&=&
\cos\phi\cos\phi' 
\hat{\gamma}_{\bf k}^{xx}
+
\sin\phi\sin\phi' 
\hat{\gamma}_{\bf k}^{yy}
\nonumber\\
&&
+
\cos\phi\sin\phi'
\hat{\gamma}_{\bf k}^{yx}
+
\cos\phi\sin\phi' 
\hat{\gamma}_{\bf k}^{xy},
\label{ramanangle}
\end{eqnarray}
where ${\bf e}_{\rm i}=(\cos\phi,\sin\phi)$
and ${\bf e}_{\rm o}=(\cos\phi',\sin\phi')$
are the directions of the {\em incoming}
and {\em outcoming} photon respectively,
and $\hat{\gamma}_{\bf k}^{ij}=
\partial^2 \hat{H}_{\bf k}/\partial k_i \partial k_j$.
Using Eq. (\ref{aham}) we obtain, close to the K point,
\begin{eqnarray}
\hat{\gamma}_{\bf k}^{xx}(\mbox{K})
&=&
-\hat{\gamma}_{\bf k}^{yy}(\mbox{K})
=
\hat{I}(\hat{\sigma}_x)/4a^2,
\\
\hat{\gamma}_{\bf k}^{xy}(\mbox{K})
&=&
\hat{\gamma}_{\bf k}^{yx}(\mbox{K})
=
-\hat{I}(\hat{\sigma}_y)/4a^2,
\end{eqnarray}
while, at the K$'$ point,
\begin{eqnarray}
\hat{\gamma}_{\bf k}^{xx}(\mbox{K}')
&=&
-\hat{\gamma}_{\bf k}^{yy}(\mbox{K}')
=
\hat{I}(\hat{\sigma}_x)/4a^2,
\\
\hat{\gamma}_{\bf k}^{xy}(\mbox{K}')
&=&
\hat{\gamma}_{\bf k}^{yx}(\mbox{K}')
=
\hat{I}(\hat{\sigma}_y)/4a^2.
\label{a4}
\end{eqnarray}

We can write the Raman (RM) response function
in the Matsubara imaginary time
for generic polarization as:
\begin{eqnarray}
\chi_{\rm RM}(\tau,\phi,\phi')
&=&
-\left\langle T_\tau
\gamma(\tau,\phi,\phi')
\gamma(\phi,\phi')
\right\rangle.
\end{eqnarray}
For unpolarized Raman scattering we have to average
over $\phi$ and $\phi'$,
$\chi_{\rm RM}(i\omega_m)=\int d\phi/(2\pi) d\phi'/(2\pi)
\chi_{\rm RM}(i\omega_m,\phi,\phi')$.
We get then
\begin{eqnarray}
\chi_{\rm RM}(i\omega_m)
&=&
\chi_{\gamma^{xx}\gamma^{xx}}(i\omega_m)
+
\chi_{\gamma^{yy}\gamma^{yy}}(i\omega_m)
\nonumber\\
&&
+
\chi_{\gamma^{xy}\gamma^{xy}}(i\omega_m)
+
\chi_{\gamma^{yx}\gamma^{yx}}(i\omega_m)
\nonumber\\
&=&
\sum_{i=x,y}
\chi_{\gamma^{ii}\gamma^{ii}}(i\omega_m)
+
\sum_{i=x,y}
\chi_{\gamma^{i\bar{i}}\gamma^{i\bar{i}}}(i\omega_m),
\end{eqnarray}
where we used the shorthand notation
$\bar{x}=y$, $\bar{y}=x$.

For the electronic Raman scattering, taking also in account
the different K and K$'$ points, these four contributes
are degenerate and we get on the real frequency axis:
\begin{eqnarray}
\chi_{\rm RM}(\omega)
&=&
4\chi_{\gamma^{xx}\gamma^{xx}}(\omega).
\end{eqnarray}
Note that, since $\hat{\gamma}_{\bf k}^{xx} \propto
\hat{I}(\hat{\sigma}_x) \propto \hat{j}_{{\bf k},x}$,
this results explicitly shows
also that the electronic Raman background
is directly related to the
electronic optical conductivity.

Let us focus now on the onset of phononic peaks within the framework
of the Fano-Rice theory.
For a generic case, for finite charge concentration $n$
and finite gap $\Delta$, we can write:
\begin{eqnarray}
\Delta\chi_{\rm RM}(\omega)
&=&
\sum_{i=x,y}\sum_{\nu\nu'}
\chi_{\gamma^{ii}\nu}(\omega)
D_{\nu\nu'}(\omega)
\chi_{\nu'\gamma^{ii}}(\omega)
\nonumber\\
&&
+\sum_{i=x,y}
\chi_{\gamma^{i\bar{i}}\nu}(\omega)
D_{\nu\nu'}(\omega)
\chi_{\nu'\gamma^{i\bar{i}}}(\omega),
\label{long}
\end{eqnarray}
where the label $\nu$ specifies at the same time the A vs S phonon
branch and the $x$ vs $y$ direction of the lattice displacement.
The notation $\bar{i}$ denotes in addition $\bar{i}=y$ if $i=x$
and  $\bar{i}=x$ if $i=y$.

Expanding explicitly the sums over $i$, $\bar{i}$, $\nu$, and $\nu'$
in Eq. (\ref{long}), as well as the summation over the different
K and K$'$ points of the Brillouin zone,
and taking into account the degeneracies
$\hat{\gamma}_{\bf k}^{xx}=-\hat{\gamma}_{\bf k}^{yy}$,
$\hat{\gamma}_{\bf k}^{xy}=\hat{\gamma}_{\bf k}^{yx}$,
$D_{{\rm S}_x{\rm S}_x}=D_{{\rm S}_y{\rm S}_y}$, 
$D_{{\rm A}_x{\rm A}_x}=D_{{\rm A}_y{\rm A}_y}$, 
$D_{{\rm A}_x{\rm S}_x}=D_{{\rm A}_y{\rm S}_y}$,
$D_{{\rm S}_x{\rm A}_x}=D_{{\rm S}_y{\rm A}_y}$, 
after few straightforward steps, we end up with
\begin{eqnarray}
\Delta\chi_{\rm RM}(\omega)
&=&
\chi_{\gamma{\rm S}}(\omega)D_{\rm SS}(\omega)
\chi_{{\rm S}^\dagger \gamma}(\omega)
\nonumber\\
&&
+\chi_{\gamma{\rm A}}(\omega)D_{\rm AA}(\omega)
\chi_{{\rm A}^\dagger \gamma}(\omega)
\nonumber\\
&&
+
\chi_{\gamma{\rm S}}(\omega)D_{\rm SA}(\omega)
\chi_{{\rm A}^\dagger \gamma}(\omega)
\nonumber\\
&&
+
\chi_{\gamma{\rm A}}(\omega)D_{\rm AS}(\omega)
\chi_{{\rm S}^\dagger \gamma}(\omega),
\label{app-rm}
\end{eqnarray}
where, due to the degeneracies of the systems,
the Raman vertex operators and the electron-phonon vertex operators
have been chosen for simplicity in Eq. (\ref{app-rm}) as
$\gamma\equiv \gamma^{xx}$, $\mbox{S}=\mbox{S}_x$, $\mbox{A}=\mbox{A}_x$.


\begin{thebibliography}{99}
\bibitem{noteeta}
For the sake of simplicity, we include here 
in a phenomenological constant term $\eta$ responsible for
all the damping processes arising from different sources of
scattering,
i.e. disorder/impurities, electron-electron and electron-lattice
interaction.
\bibitem{noteTB}
We remind here that the vanishing of $\chi_{j{\rm A}}$ in the undoped
bilayer graphene,
and hence of the phonon activity of this mode,
is true if additional terms in the tight-binding
Hamiltonian, such as the crystal field or the
$\gamma_3$, $\gamma_4$
hopping parameters are neglected. A more
detailed discussion of this issue is provided in Section \ref{s-TB}
\end{thebibliography}
\end{document}